\documentclass[fleqn,usenatbib]{mnras}

\usepackage{newtxtext,newtxmath}

\usepackage[T1]{fontenc}
\usepackage{ae,aecompl}

\usepackage{graphicx}	
\usepackage{amsmath}	
\usepackage{amssymb}	

\usepackage[compatibility=false]{caption}
\usepackage{subcaption}

\title[The accretion rate - dust mass correlation]{A dusty origin for the correlation between protoplanetary disc accretion rates and dust masses}

\author[A. D. Sellek et al.]{
Andrew D. Sellek,$^{1}$\thanks{E-mail: ads79@cam.ac.uk}
Richard A. Booth,$^{1}$
Cathie J. Clarke$^{1}$
\\
$^{1}$Institute of Astronomy, University of Cambridge, Madingley Road, Cambridge CB3 0HA, UK\\
}

\date{Accepted 2020 August 17. Received 2020 August 14; in original form 2020 July 3}

\pubyear{2020}

\begin{document}
\label{firstpage}
\pagerange{\pageref{firstpage}--\pageref{lastpage}}
\maketitle

\begin{abstract}
Recent observations have uncovered a correlation between the accretion rates (measured from the UV continuum excess) of protoplanetary discs and their masses inferred from observations of the sub-mm continuum.
While viscous evolution models predict such a correlation, the predicted values are in tension with data obtained from the Lupus and Upper Scorpius star forming regions; for example, they underpredict the scatter in accretion rates, particularly in older regions. Here we argue that since the sub-mm observations trace the discs’ dust, by explicitly modelling the dust grain growth, evolution, and emission, we can better understand the correlation.
We show that for turbulent viscosities with $\alpha \lesssim 10^{-3}$, the depletion of dust from the disc due to radial drift means we can reproduce the range of masses and accretion rates seen in the Lupus and Upper Sco datasets. One consequence of this model is that the upper locus of accretion rates at a given dust mass does not evolve with the age of the region.
Moreover, we find that internal photoevaporation is necessary to produce the lowest accretion rates observed. In order to replicate the correct dust masses at the time of disc dispersal, we favour relatively low photoevaporation rates $\lesssim 10^{-9}~\mathrm{M_{\sun}}~\mathrm{yr^{-1}}$ for most sources but cannot discriminate between EUV or X-ray driven winds.
A limited number of sources, particularly in Lupus, are shown to have higher masses than predicted by our models which may be evidence for variations in the properties of the dust or dust trapping induced in substructures.
\end{abstract}




\section{Introduction}
It is now well established that young stars are frequently surrounded by discs of gas and dust, which are generally regarded as the sites of planet formation. Studying these environments allows us to constrain the processes in the disc gas and dust that both promote and halt the assembly of the diverse planetary systems that we observe.

One such process, the accretion of the material onto the central star, has been detected in most of these protoplanetary discs by measuring either the UV continuum luminosity excess \citep{Alcala_2014,Alcala_2017}, or line luminosities from H$\alpha$ \citep{Fedele_2010} or CIV \citep{Alcala_2019}. This not only shows that protoplanetary discs act as accretion discs \citep[a suggestion that goes back to][]{Lynden-Bell_Pringle_1974}, but the measurements have been used for many years to provide important insights into the evolution of the circumstellar environment, for example constraining the initial conditions of disc formation \citep{Dullemond_2006,Alexander_2006} or the rates of disc dispersal \citep{Ercolano_2014}.

A key question in current studies of protoplanetary discs is how the accretion rates relate to the masses of the discs that supply them, and the origin of this relation.
While it is still not entirely clear what drives the accretion \citep[e.g.][]{Rafikov_2017}, it is clear that it requires the loss of angular momentum from the material that accretes.
The angular momentum may be removed from the disc entirely by magnetic torques associated with a hydromagnetic wind, or redistributed within the disc by an effective viscosity.
To gain further insight, we may use the accretion rates $\dot{M}_{\rm acc}$ and masses $M_{\rm disc}$ together to calculate the \textit{accretion timescale}\footnote{Also called the `disc age' by \citet{Jones_2012} or 'disc lifetime' by \citet{Lodato_2017}}
\begin{equation}
    t_{\rm acc} = \frac{M_{\rm disc}}{\dot{M}_{\rm acc}}
    \label{eq:tacc_M}
    .
\end{equation}

The accretion timescale may be used to constrain the strength of the effective viscosity, or other process, that drives accretion. 
Within the viscous framework, the effective viscosity $\nu$ is most commonly parametrized in terms of the \citet{Shakura-Sunyaev_1973} $\alpha$ model. Physically, the dimensionless $\alpha$ parameter represents the strength of the viscosity relative to the maximum turbulence the disc could support.
However we can also interpret $\alpha$, for a given disc geometry, as a measure of the "viscous timescale" - the time for the material at a certain radius to reach the star relative to its orbital time (regardless of the actual mechanism responsible for the angular momentum evolution).
In these terms, the accretion timescale, which measures the period over which the whole disc could accrete, is measuring the viscous timescale for the outer radius of the disc.
This has lead to measurements of $\alpha$ for individual discs with typical values of $\lesssim10^{-2}$ \citep{Rafikov_2017,Ansdell_2018}. Such values are roughly consistent with the evolution of accretion rates in Taurus which imply $\alpha \sim 10^{-2}$ \citep{Hartmann_1998}.

Viscous evolution models predict how the ratio of disc mass to accretion rate - the accretion timescale - varies as a function of time.
In order to conserve angular momentum, a viscous disc must balance the inward accretion with an outward motion of material known as viscous spreading. This moves the outer radius outwards; since typically the timescale for viscous evolution is an increasing function of radius, this increases the accretion timescale.
The longer a disc has had to evolve, the larger a radius can have been reached by viscous spreading. Therefore, after a few multiples of the initial accretion timescale, the outer radius is wherever the viscous timescale is roughly the system age.
Thus, in the case of a power law dependence of viscosity on radius, viscous models predict a self-similar evolution in which the accretion timescale depends only on time and is independent of the disc initial conditions.
Hence, at a (sufficiently large) given time, viscous evolution models for discs predict a linear relationship between the accretion rate and disc mass.

The full expression for the accretion timescale (which correctly reduces to $t_{\rm acc} \sim t$ for $t \gg t_\nu$) for a power law dependence of viscosity on radius ($\nu \propto R^\gamma$)
is \citep[e.g.][]{Lodato_2017}:
\begin{equation}
    t_{\rm acc} = 2(2-\gamma)(t+t_\nu)
    \label{eq:tacc_t}
    ,
\end{equation}
where $t$ is the system age and $t_\nu$ is the initial viscous timescale.
This relationship between the accretion rate and disc mass can also be expressed in terms of the `dimensionless accretion parameter' $\eta = \frac{t}{t_{\rm acc}}$ \citep{Rosotti_2017}.

\citet{Manara_2016} found the first evidence, in Lupus, of a correlation between the accretion rates \citep[measured by][]{Alcala_2014,Alcala_2017} and the disc masses \citep[measured from the dust emission by][]{Ansdell_2016} with a power law slope of $0.7 \pm 0.2$, consistent with a simple linear relationship as expected from the viscous theory above.
Assuming a dimensionless accretion parameter of unity, they found that 60 per cent of discs were consistent with an accretion timescale of $1-3~\mathrm{Myr}$, the age of the Lupus region. However, several sources had accretion timescales that were either too long (i.e. too low an accretion rate for their mass), or too short (too high an accretion rate for their mass) that is there was too large a spread in $t_{\rm acc}$.
Equivalently, to be explained by the self-similar evolution some discs had to be too old or too young for the age of the Lupus region.

The correlation between accretion rate at disc mass at a given age is expected to become tighter with age.
Equation \ref{eq:tacc_t} implies that for a region with a given age, discs should show a spread in $t_{\rm acc}$ values owing to there being a range of initial viscous times $t_\nu$, but the relative scatter should decrease with age \citet{Lodato_2017}.
The region Chamaeleon I does show a very similar range of disc masses and accretion rates to Lupus, as befits its similar age \citep{Mulders_2017}.
On the other hand the Upper Scorpius region, which is $5-10~\mathrm{Myr}$ old \citep{Pecaut_2016,David_2019} should show lower accretion rates at a given mass, as well as a tighter spread in $t_{\rm acc}$. However, \citet{Manara_2020} found that the median accretion rates and spreads are not dissimilar to the younger regions, resulting in a number of discs which could only be explained by Equation \ref{eq:tacc_t} if they had very young ages (i.e. they exhibit very high accretion rates for their masses).

Discs that appear too \textit{old} (i.e. have $t_{\rm acc} \gg t$) can be readily explained if the initial viscous timescale is sufficiently long compared to the system age. Then Equation \ref{eq:tacc_t} shows that $t_{\rm acc}$ is instead a measure of this initial timescale \citep{Rosotti_2017,Lodato_2017}.
However, for the discs in the Lupus sample, which have $\langle t_{\rm acc} \rangle \approx 2.5~\mathrm{Myr}$ and $\langle t \rangle \approx 1.6~\mathrm{Myr}$, \citet{Lodato_2017} found this was not the case, and instead used $t_{\rm acc}$ to place a further constraint on the viscosity: Equation \ref{eq:tacc_t} implies that for $t_\nu>0$, $\gamma > 2 - \frac{1}{2\eta}$, and thus a viscous model is only consistent with these data if $\gamma>1.2$. Equivalently, if $\gamma<1.2$, as in many models which assume a constant $\alpha$ throughout the disc, the observed $\langle t_{\rm acc} \rangle$ means the discs appear too \textit{young} to agree with the cluster age.
As noted above, this discrepancy is even more extreme for Upper Sco.

Several potential complications to the viscous model - such as dead zones, planet formation and internal photoevaporation - have been considered but tend to reduce accretion rates, and hence also increase $t_{\rm acc}$ \citep{Jones_2012,Rosotti_2017}.
External photoevaporation does reduce $t_{\rm acc}$ \citep{Rosotti_2017} and may be relevant in Upper Sco \citep{Trapman_2020}, but is not thought to be a significant influence in Lupus: for example the trend observed between dust disc radii and sub-mm flux \citep{Tripathi_2017,Andrews_2018} precludes strongly photoevaporating discs \citep{Sellek_2020}.

So far, all of this has been predicated on masses inferred from dust continuum observations with the canonical assumption that the gas-to-dust ratio retains its primordial value of 100.
\citet{Manara_2016} found no correlation between accretion rates and gas masses estimated from CO emission lines.
Several studies have thrown into doubt the accuracy of using the masses of CO isotopologues to trace the total mass \citep[e.g.][]{Miotello_2017,Bosman_2018,Powell_2019}, suggesting that CO-based data are not a reliable measure.
However this prompts the question as to whether, as \citet{Manara_2016} have argued, the relative success of the dust masses in producing agreement with the viscous predictions (despite the issues highlighted above) is a vindication of deriving disc masses from dust emission under standard assumptions.
\citet{Mulders_2017} found that agreement with data in Lupus and Chamaeleon I could be further improved by introducing an ad hoc elevation and scatter in the assumed gas-to-dust ratio.
However, to date there have been no attempts to interpret this data using models for grain growth and evolution \citep[although such models have been successfully used to explain the correlation between the disc sub-mm fluxes and dust disc radii][]{Rosotti_2019}.

Conducting the dust modelling is important as theoretical models of dust evolution predict a much more complex mapping between dust emission and total disc mass than is generally assumed.
The gas-to-dust ratio may deviate significantly from 100 due to effects such as radial drift, grain growth, dust trapping and internal or external photoevaporation \citep[e.g.][]{Takeuchi_2005,Alexander_2007,Birnstiel_2012,Sellek_2020}. Moreover, models predict opacities and temperatures that vary with properties such as grain size, composition and location, rather than the single values used by \citet[e.g.][]{Ansdell_2016} in estimating dust masses from sub-millimetre fluxes. Discrepancies between the true and observed masses could also arise due to optically thick emission \citep{Galvan-Madrid_2018}. 
There is also observational evidence for this paradigm: by modelling the extent of the dust emission at different wavelengths and using dynamical arguments, \citet{Powell_2019} derived mass estimates independent of any tracer-to-total mass conversion which implied gas-to-dust ratios of $10^{3}-10^{4}$.

Given this potential variation in dust-to-gas ratio and opacity, it is somewhat surprising that that the dust masses show a meaningful correlation with the accretion rates.
This suggests a re-evaluation of the interpretation of dust masses is necessary and may be another way of redressing the tension between the apparently young, widely spread, ages of observed discs and the older, more tightly correlated, predictions of viscous evolution models.
The observed relationships in Lupus and Upper Sco thus allow us to test theoretical models of dust evolution by seeing if the models can replicate the relationships across a range of ages.

In this work, we focus on understanding which parts of the $\dot{M}_{\rm acc}-M_{\rm dust}$ plane dust growth and drift models can reach and whether these are compatible with the observations, rather than reproducing the exact trends with population synthesis.
We use the dust model of \citet{Birnstiel_2012} to compute the evolution of the dust distribution in viscously accreting discs.
From these models, we calculate a sub-mm flux density from which we estimate an "observers' equivalent dust mass" in order to account for any differences between the true and observationally-inferred masses and hence study the evolution of discs in the same observational plane as the survey data. The disc model and details of this conversion are set out in Section \ref{sec:model}.
In Section \ref{sec:results}, we present results to illustrate the improved ability of a model accounting for dust evolution, allowing for a range of initial disc parameters, to reproduce observations of the accretion rate and disc mass in Lupus (representing a younger region) and Upper Sco (representing an older region).
In the following sections we discuss caveats and remedies to the remaining outliers:
in Section \ref{sec:photoevaporation}, the impact of both EUV and X-ray internal photoevaporation models on these results;
in Section \ref{sec:parameters}, varying the model parameters such as the properties of the dust evolution or the stellar mass;
and in Section \ref{sec:other}, the effects of binaries or dust trapping.
Finally, we summarise our conclusions in Section \ref{sec:conclusions}.

\section{Model Description}
\label{sec:model}
In this work we use a version of the model of \citet{Booth_2017}, which solves the viscous diffusion equation for a \citet{Shakura-Sunyaev_1973} $\alpha$ viscosity model and includes a two-population dust growth model following \citet{Birnstiel_2012}. We also add photoevaporation into the model; an almost identical model of viscous evolution, dust and photoevaporation was first used by \citet{Ercolano_2017}. In all cases, the evolution equations were solved on a grid with 5000 cells equispaced in $R^{1/2}$ between $0.025~\mathrm{au}$ and $10000~\mathrm{au}$.

\subsection{Gas Evolution}
\label{sec:gas}
The viscous diffusion equation for the evolution of the surface density $\Sigma(R,t)$ of the gas in a disc is
\begin{equation}
    \frac{\partial \Sigma}{\partial t} = \frac{1}{3 R} \frac{\partial}{\partial R} \left(R^{1/2} \frac{\partial}{\partial R} \left( \nu R^{1/2} \Sigma \right) \right) - \dot{\Sigma}_{\rm photo}
    ,
    \label{eq:gasevol}
\end{equation}
where $\dot{\Sigma}_{\rm photo}$ represents the mass loss due to internal photoevaporation.
In our basic model we do not include any photoevaporation. However, in Sections \ref{sec:photoevaporation} and \ref{sec:other}, we introduce models that use either the EUV prescriptions parametrized by \citet[][Equations A1-A5]{Alexander_2007} or the X-ray prescriptions derived by \citet[][Equations 2-5]{Picogna_2019} for $\dot{\Sigma}_{\rm photo}$; the latter are a recent update to the commonly-used prescriptions presented by \citet{Owen_2012}. Once photoevaporation opens a gap, and the column density to the outer edge of this gap is $<10^{22}~\mathrm{cm}^{-2}$, we switch from the `Primordial disc' to the `Inner-Hole disc' for $R \geq R_{\rm hole}$ while continuing to use the `Primordial' profile at smaller radii to clear the remaining inner disc. The fit to the total mass-loss rate in the primordial case quoted by \citet{Picogna_2019} is as follows, with an inner-hole disc sustaining a rate $1.12$ times higher:
\begin{equation}
    \log_{10}(\dot{M})) = -7.2580 -2.7326 \exp\left[ -\frac{(\ln(\log_{10}(L_{\rm X}))-3.3307)^2}{2.9868\times10^{-3}} \right],
    \label{eq:Picogna_Mdot}
\end{equation}
where $L_{\rm X}$ is the X-ray luminosity in $\mathrm{erg~s^{-1}}$. We take this expression at face value and assume all dependence on the star is through $L_{\rm X}$, as the explicit effect of stellar mass is weak \citep{Owen_2012}. However, we rescale the \textit{profiles} provided by \citet{Picogna_2019} radially by stellar mass according to Equations \ref{eq:Picogna_radius_scale_P} and \ref{eq:Picogna_radius_scale_TD} \citep[c.f. Equations B3, B6 of][]{Owen_2012} to account for the gravitational effect of masses different to their fiducial $0.7 \mathrm{M_{\sun}}$.

\begin{equation}
    R' = R \left( \frac{M_*}{0.7 \mathrm{M_{\sun}}} \right)^{-1},
    \label{eq:Picogna_radius_scale_P}
\end{equation}
\begin{equation}
    x' = (R-R_{\rm hole}) \left( \frac{M_*}{0.7 \mathrm{M_{\sun}}} \right)^{-1}.
    \label{eq:Picogna_radius_scale_TD}
\end{equation}

We choose not to include external photoevaporation as the typical FUV radiation fields are only $\sim 4~G_0${\footnote{The Habing unit, ${\rm G_0}$, which is $1.6 \times 10^{-3}$ erg cm$^{-2}$ over the energy range $6-13.6$ eV \citep{Habing_1968}, is defined such that the mean interstellar FUV field is $1~{\rm G_0}$}}
in Lupus \citep{Cleeves_2016} \citep[although this might have an effect for sufficiently extended discs such as IM Lup,][]{Haworth_2017}. The typical FUV field strengths are likely higher in Upper Sco \citep[][]{Trapman_2020}, but for a more direct comparison between Upper Sco and Lupus we neglect their effect.

The $\alpha$ viscosity model of \citet{Shakura-Sunyaev_1973} assumes that the viscosity $\nu$ can be parametrized as
\begin{equation}
    \nu = \alpha c_{\rm S} H
    ,
\end{equation}
where $c_{\rm S}$ is the sound speed and $H=c_{\rm S}/\Omega$ is the scale height of the disc.
This means that
\begin{equation}
    \nu = \alpha \frac{c_{\rm S}^2}{\Omega} \propto \alpha R^{3/2} T
    .
\end{equation}
Assuming that the disc is heated by the dust, which in turn is heated by stellar irradiation, we expect the temperature to scale with radius as $T\propto R^{-1/2}$ \citep{Kenyon_1987,Kenyon_1995,Chiang_1997,Andrews_2005}.
Although some authors choose to consider a non-constant $\alpha$ \citep[e.g.][where $\alpha \propto R^{1/2}$ such that $\nu \propto R^{3/2}$]{Lodato_2017}, in this work we choose a constant $\alpha=10^{-3}$; under the assumption of a constant $\alpha$, $\nu \propto R$. This value of $\alpha$ corresponds to the case for which dust drift models are better able to reproduce the disc flux-size relationships \citep{Rosotti_2019,Sellek_2020}.
The temperature is set by imposing an aspect ratio (the ratio of the scale height over the radius) of $0.033$ at $1~\mathrm{au}$.

For a viscosity law $\nu \propto R^\gamma$, \citet{Lynden-Bell_Pringle_1974} showed that there is a generally-attracting similarity solution to Equation \ref{eq:gasevol} of the form
\begin{equation}
    \Sigma(R,t) = \frac{M_{\rm disc,0}}{2\pi R_{\rm C}^2} (2-\gamma) \left(\frac{R}{R_{\rm C}}\right)^{-\gamma} \left(1+\frac{t}{t_\nu}\right)^{-\eta}
    \exp\left(- \frac{(R/R_{\rm C})^{2-\gamma}}{\left(1+\frac{t}{t_\nu}\right)}\right)
    ,
    \label{eq:LBPprofile}
\end{equation}
where $\eta = (5/2-\gamma)/(2-\gamma)$.
These solutions have a `accretion timescale' given by Equation \ref{eq:tacc_t}, where the initial viscous timescale is defined as follows, with a typical value for our parameters shown
\begin{align}
    t_\nu
    &= \frac{1}{3(2-\gamma)^2}\frac{R_{\rm C}^2}{\nu(R_{\rm C})}
    \label{eq:tvisc} \\
    &= 4.89~\mathrm{Myr}~\left(\frac{R_{\rm C}}{100~\mathrm{au}}\right) \left(\frac{\alpha}{10^{-3}}\right)^{-1} \left(\frac{h_0}{0.033}\right)^{-2} \left(\frac{M_*}{\mathrm{M_{\sun}}}\right)^{-1/2} \nonumber
\end{align}
In line with our constant $\alpha$, we choose the $\gamma=1$ solution for our initial condition, and explore the same observationally-motivated values for the initial scale radii $R_{\rm C}=\{10,30,100\}~\mathrm{au}$ and initial masses $M_{\rm disc,0}=\{1,3,10,30,100\}~\mathrm{M_{\rm J}}$ (where $\mathrm{M_{\rm J}}$ is the mass of Jupiter) as in \citet{Sellek_2020}.

\subsection{Dust Evolution}
\label{sec:dust}
Key to this work (as compared to other studies such as the recent \citet{Appelgren_2020} that use a single constant dust grain size) is that we use the two population model of \citet{Birnstiel_2012} which allows the maximum grain size to vary with local conditions.
It is important to know the size of the largest grains for two reasons. Firstly, the size affects the dynamics of the dust and hence the evolution of the dust budget. Secondly, for comparison to observations, it is the largest grains that dominate the emissivity at mm wavelengths.

The dynamics of the dust are affected by drag and thus governed by the Stokes number, a dimensionless measure of the stopping time, which in the Epstein regime is given by \citep{Birnstiel_2012}
\begin{equation}
    St = \frac{\pi}{2} \frac{a \rho_{\rm s}}{\Sigma_{\rm gas}}
    \label{eq:Stokes}
    .
\end{equation}
To zeroth order, the dust is advected with the gas but as the dust size (Stokes number) increases it becomes increasingly decoupled from the gas as the stopping times become long. Moreover, since the pressure support leads to the gas, but not the dust, experiencing sub-Keplerian rotation, a headwind results, ultimately leading to the radial drift of dust due to drag-induced torques \citep{Whipple_1973,Weidenschilling_1977}.

\citet{Birnstiel_2012} showed that the dust evolution could be well-reproduced by a simple model considering two populations: a small population of well-coupled monomer grains, that can grow into larger dust, and a large population of dynamically less well-coupled grains. At each radius in the disc the maximum size $a$ of the dust grain distribution (as represented by this large population) is allowed to grow as
\begin{equation}
    \frac{a}{\dot{a}} = \frac{f_{\rm grow}}{\epsilon \Omega}
    ,
\end{equation}
where $f_{\rm grow}$ is an extra factor used by e.g. \citet{Ormel_2017,Booth_2020} which is an (inverse) measure of the fraction of dust grain collisions that lead to growth. Growth stops once the dust size reaches the lower of the fragmentation-limited or drift-limited regime \citep{Birnstiel_2012}:
\begin{equation}
    a_{\rm frag} = 0.37 \frac{2}{3\pi} \frac{\Sigma_{\rm gas}}{\rho_{\rm s} \alpha} \frac{u_f^2}{c_{\rm S}^2}
    \label{eq:afrag}
\end{equation}
\begin{equation}
    a_{\rm drift} = 0.55 \frac{2}{\pi} \frac{\Sigma_{\rm dust}}{\rho_{\rm s} f_{\rm grow}} \frac{v_{\rm K}^2}{c_{\rm S}^2} \left| \frac{\rm{d} \ln P}{\rm{d} \ln R} \right|^{-1}
   \label{eq:adrift}
\end{equation}
Numerical values assumed for dust properties are listed in Table \ref{tab:dust_model}.

To implement the evolution, we use the model of \cite{Booth_2017} which uses the relative dust-gas velocities to update the dust mass fractions based on \citet{Laibe_2014}. However, we have updated their model to use the relative velocities given by the following formulae from \citet{Dipierro_2018}, which include more completely the effects of dust feedback and imperfect dust-gas coupling (although the differences should only be significant when $St \gg 1$).
The azimuthal and radial components of the gas velocity are
\begin{align}
    v_{r,\rm gas} &=
    - \frac{\lambda_1}{(1+\lambda_0)^2+\lambda_1^2} v_P
    + \frac{1+\lambda_0}{(1+\lambda_0)^2+\lambda_1^2} v_{\rm visc} \\
    v_{\phi,\rm gas} &=
    \frac{1}{2} \frac{1+\lambda_0}{(1+\lambda_0)^2+\lambda_1^2} v_P
    +\frac{1}{2} \frac{\lambda_1}{(1+\lambda_0)^2+\lambda_1^2} v_{\rm visc}
    ,
\end{align}
where $\lambda_k = \sum_i \frac{\epsilon_i}{1-\epsilon} \frac{St_i^k}{1+St_i^2}$. $\epsilon_i$ is the fraction of mass in dust species $i$ (such that the total dust mass fraction is $\epsilon = \sum_i \epsilon_i$) - with two species in the two population model.
$v_{\rm visc} = \frac{3}{\Sigma_G} \frac{\partial}{\partial R} \left( \nu R^{1/2} \Sigma_G\right)$ is the velocity induced by viscous torques and $v_P = \frac{c_{\rm S}^2}{v_{\rm K}} \frac{\rm{d} \ln P}{\rm{d} \ln R}$.
From these, the radial velocity of dust species $i$ relative to the gas may be calculated as
\begin{equation}
    \Delta v_{r,i} = \frac{2 v_{\phi,\rm gas} St_i - v_{r,\rm gas} St_i^2}{1+St_i^2}
    .
\end{equation}

\begin{table*}
    \centering
    \caption{Numerical Values of Gas Model, Dust Model and Photoevaporation Model Parameters in the Basic Model. Variations to these examined in Sections \ref{sec:photoevaporation} \& \ref{sec:parameters} are included in round brackets.}
    \label{tab:dust_model}
    \begin{tabular}{r|l|c}
        \hline
        Parameter   & Value  & Reference \\
        \hline
        Viscosity ($\alpha$)                    & $10^{-3}$ ($3\times10^{-4},3\times10^{-3}$)
        & \citet{Rosotti_2019,Sellek_2020} \\
        Aspect Ratio at $1~\mathrm{au}$ ($h_0$) & $0.033$ ($0.025$)
        & \citet{Rosotti_2019} \\
        Initial Mass ($M_{\rm disc,0}$)                & $\{1,3,10,30,100\}~\mathrm{M_{\rm J}}$
        & - \\
        Initial Scale Radius ($R_{\rm C}$)            & $\{10,30,100\}~\mathrm{au}$
        & - \\
        Stellar Mass ($M_*$)                    & $1.0~\mathrm{M_{\sun}}$ ($0.1~\mathrm{M_{\sun}}$)
        & - \\
        \hline
    Internal Dust Density ($\rho_{\rm s}$)        & $1.0~\mathrm{g~cm^{-3}}$ 
        & \citet{Tazzari_2016,Pollack_1994} \\
        Fragmentation velocity ($u_f$)          & $10~\mathrm{m~s^{-1}}$ ($1~\mathrm{m~s^{-1}}$)
        & \citet{Gundlach_2015} \\
        Initial/Minimum size ($a_0$)            & $0.1 ~ \mathrm{\mu m}$
        & - \\
        Initial dust-to-gas ratio ($\epsilon$)  & $0.01$
        & \citet{Bohlin_1978} \\
        Growth factor ($f_{\rm grow}$)          & 1 (10)
        & \citet{Birnstiel_2012}; \citep{Booth_2020} \\
        \hline
        X-ray Luminosity ($L_{\rm X}$)                & $0~\mathrm{erg~s^{-1}}$ ($5\times10^{28},10^{30},10^{31}~\mathrm{erg~s^{-1}}$)
        & \citep{Preibisch_2005,Gudel_2007} \\
        EUV Photon Flux ($\Phi$)       & $0~\mathrm{s^{-1}}$ ($10^{42}~\mathrm{s^{-1}}$)
        & \citep{Pascucci_2009} \\
        \hline
    \end{tabular}
\end{table*}

\subsection{Calculation of Flux and Observers' Equivalent Dust Mass}
\label{sec:obsmasses}
Although dust masses are easily retrieved directly from the model by integrating the surface density of the dust over the disc area, the observational data to which we wish to compare is based on masses inferred from the sub-mm flux densities \citep{Ansdell_2016}. To ensure we compare like-for-like, we thus calculate an "observers' equivalent dust mass" in the following way \citep[which is the same as the calculation of the "synthetic dust mass" of][]{Pinilla_2020}.

The sub-mm flux density from the disc is first calculated assuming emission, neglecting scattering, from a vertically isothermal, face-on\footnote{So long as the disc is optically thin, which is mostly the case here, $F_\nu$ is not sensitive to the inclination unless the discs are close to edge on.} disc as \citep[e.g.][]{Tanaka_2005,Tazzari_2016,Pinilla_2020}
\begin{equation}
    F_\nu = \frac{1}{d^2} \int_{R_{\rm in}}^{R_{\rm out}} 2\pi R B_\nu(T) (1-e^{-\tau_\nu}) dR
    ,
    \label{eq:flux}
\end{equation}
where $B_\nu(T)$ is the Planck function at temperature $T(R)$, $\tau_\nu = \kappa_\nu(a) \Sigma_{\rm dust}$ is the optical depth and $d$ is the distance to the disc. To calculate the fluxes we use opacities $\kappa_\nu(a)$ which are calculated for the grain size $a_{\rm max}(R)$ using opacity tables appropriate for compact grains of the composition used by \citet{Tazzari_2016,Pollack_1994} with grain size distribution $n(a) \propto a^{-3.5}$. The temperature used is the gas temperature described in \ref{sec:gas}:
\begin{equation}
    T(R) = 279 \left(\frac{R}{1~\mathrm{au}}\right)^{-0.5}  \mathrm{K}.
    \label{eq:T_R}
\end{equation}

For optically thin ($\tau_\nu<<1$) dust emission at constant temperature $T_{\rm dust}$ and opacity $\kappa_\nu$, Equation \ref{eq:flux} reduces to
\begin{equation}
    F_\nu = \frac{B_\nu(T_{\rm dust}) \kappa_\nu}{d^2} \int_{R_{\rm in}}^{R_{\rm out}} 2\pi R \Sigma_{\rm dust} dR
    .
\end{equation}
Recognising the integral as the dust mass and rearranging, this becomes
\begin{equation}
    M_{\rm d, obs} = \frac{F_\nu d^2}{B_\nu(T_{\rm dust}) \kappa_\nu}
    \label{eq:Mdobs}
    ,
\end{equation}
the relationship derived by \citet{Hildebrand_1983}. This is the simplest, and most common, method used to estimate the dust mass, including by \citet{Ansdell_2016} and \citet{Barenfeld_2016} and hence \citet{Manara_2016} and \citet{Manara_2020} for the data sets to which we compare here. Thus we also use this to estimate the dust mass from the calculated sub-mm flux densities.

Although other observational studies do use more sophisticated prescriptions for $T_{\rm dust}$ that depend on the stellar luminosity and/or the flux profile of the disc \citep[e.g.][]{Andrews_2013,Barenfeld_2016,Tripathi_2017}, for consistency with \citet{Ansdell_2016}, we use a single dust temperature $T_{\rm dust} = 20~\mathrm{K}$ and $\kappa_\nu = 10~ \mathrm{cm^2~g^{-1}} \left(\frac{\nu}{1000~\mathrm{GHz}}\right)$.
To avoid making any assumptions about the stellar luminosity, we rescale the masses from \citet{Manara_2020} to use the same dust temperature and opacity as \cite{Ansdell_2016}.
The calculations are performed at $890~\mathrm{\mu m}$ when comparing to Lupus or $880~\mathrm{\mu m}$ when comparing to Upper Sco, though this makes negligible difference.

\subsection{Summary of Nomenclature}
\label{sec:nomenclature}
For what concerns the masses, the "true dust mass" is the actual mass in dust that our models predict a disc would contain. The "observers' equivalent dust mass" ($M_{\rm d, obs}$) is the mass in dust that an observer would infer from the same model, given the fluxes that we calculate. The details of this calculation are given in Section \ref{sec:obsmasses}. The "observers' equivalent disc mass" is the total mass that an observer would infer given standard assumptions i.e. $100$ times the "observers' equivalent dust mass".

Throughout this work we use "system age" $t$ to refer to the true age of the star and disc. The "accretion timescale" $t_{\rm acc}$ is the ratio of the observer's equivalent disc mass to the accretion rate. For the $\gamma=1$, constant $\alpha$, viscous models we employ, $t_{\rm acc} = 2 (t+t_\nu)$. Hence at late times, when $t \gg t_\nu$, $t_{\rm acc} \sim 2t$. Thus, we use the term "inferred viscous age" to refer to the age as estimated from the accretion timescale under this viscous model ($t_{\rm acc} / 2$).

\section{Basic Dust Evolution Model without Photoevaporation}
\label{sec:results}
\begin{figure*}
    \centering
	\includegraphics[width=\linewidth]{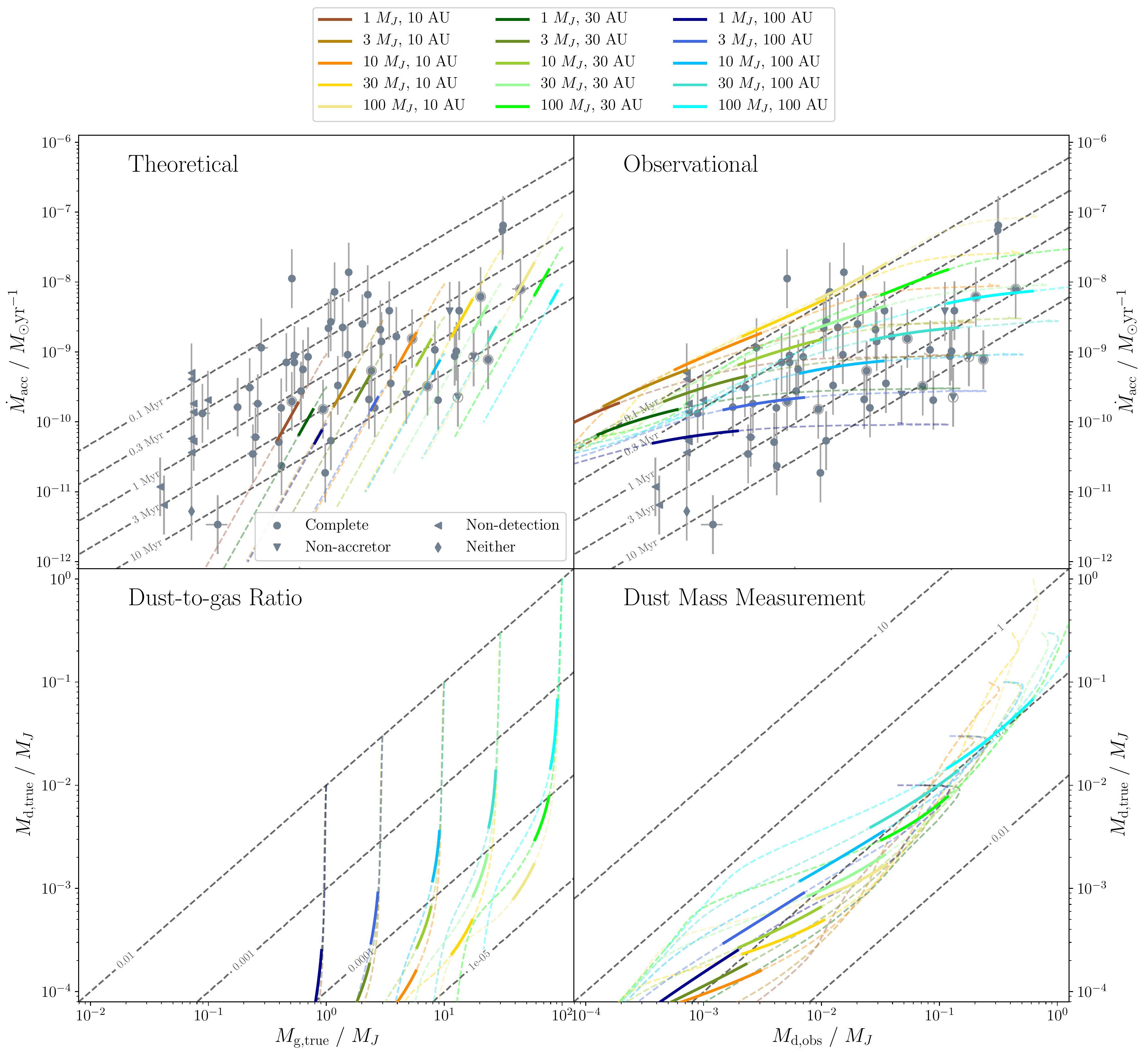}
    \caption{Model tracks using a dust evolution model are indicated by coloured dashed/solid lines with the solid section corresponding to $1-3$ Myr.
    Upper left: the `theoretical' plane shows the accretion rate against the gas mass; dashed grey lines show the expected location at the labelled time in the limit where $t \gg t_\nu$ (i.e. the labels represent the inferred viscous age).
    Lower left: the plane of true dust masses against gas masses; dashed grey lines show the dust-to-gas ratio.
    Lower right: the plane of true dust masses against observers' equivalent dust mass; dashed grey lines show the ratio between the two.
    Upper right: the `observational' plane of accretion rate against the observers' equivalent dust mass; the dashed grey lines are labelled with the inferred viscous age assuming a gas-to-dust ratio of 100.
    The grey points in the upper panels show the accretion rate \citep{Alcala_2014,Alcala_2017} and inferred gas mass (assuming a gas-to-dust ratio of 100) or dust mass \citep{Ansdell_2016} for the discs in Lupus, with the marker shape indicating whether the object has complete data, is a non-accretor, a non-detection or neither an accretor nor a detection, and transition discs drawn enclosed by a lighter ring.
    }
    \label{fig:4panels_nope}
\end{figure*}

We first present the results of the evolution of 15 disc models spanning a grid of initial radii and masses and without including any internal photoevaporation mechanism.
These models are shown as the coloured tracks in each of the panels of Figure \ref{fig:4panels_nope} and the interpretation discussed in the following sub sections.
Between $1$ and $3~\mathrm{Myr}$ the tracks are plotted with thick, solid, lines. Outside of these times, the tracks are plotted with fainter, dashed, lines of the same colour. The length of the solid section thus indicates how rapidly the disc evolves within this age window, to give an indication of the amount of time the disc spends at its location in the plane of the two quantities defining the panel. The same models are shown in Figure \ref{fig:nope_old} with the solid sections corresponding to disc locations at ages between $5$ and $10~\mathrm{Myr}$.

\subsection{Without Accounting for Dust}
\label{sec:Lupus}

By eliminating time from the evolution of the disc mass and accretion rate \citet{Lodato_2017} showed that while a population of discs obeying the similarity solution of \cite{Lynden-Bell_Pringle_1974} follows a shallow `isochrone', individual discs should evolve along steep trajectories given by
\begin{equation}
    \dot{M}_{\rm acc} = \frac{1}{2(2-\gamma)} \frac{M_{\rm disc,0}}{t_\nu} \left( \frac{M_{\rm disc}}{M_{\rm disc,0}} \right)^{5-2\gamma}
    .
    \label{eq:track}
\end{equation}

The upper left `Theoretical' panel of Figure \ref{fig:4panels_nope} shows the accretion rate against the gas mass of the disc. Each model evolves from top-right towards bottom-left along a power law track corresponding to $\dot{M}_{\rm acc} \propto M_{\rm disc}^3$ - as would be predicted from Equation \ref{eq:track} for $\gamma=1$.

As detailed in Section \ref{sec:nomenclature}, from Equation \ref{eq:tacc_t}, we see that for $\gamma=1$, the accretion timescale $t_{\rm acc}=2(t+t_\nu)$; in line with this, we use `inferred viscous age' to refer to $t_{\rm acc}/2$. Thus, in the panel we also indicate lines where $M_{\rm disc}/\dot{M}_{\rm acc}=2t$, for several values of $t$.
However, at a given time, a disc's inferred viscous age is older by a factor of $1+t_\nu/t$ than the true system age, due to the finite value of the initial viscous time $t_\nu$. This means that, particularly at young ages, the models are located slightly below and to the right of the line where their inferred viscous ages are equal to the true system ages. The solid sections of the tracks in the upper left panel of Figure \ref{fig:4panels_nope} illustrate the models evolving to longer $t_{\rm acc}$ over time as expected, while their endpoints clearly show this small offset from the indicated contours of inferred viscous age.

Moreover, since $t_\nu \propto R_{\rm C}$ (Equation \ref{eq:tvisc}), a degree of scatter in $R_{\rm C}$ leads to a spread in $t_{\rm acc}$ at a given time, (i.e. perpendicular to the linear trend between $\dot{M}_{\rm acc}$ and $M_{\rm disc}$). This manifests in the displacement of the solid tracks with different $R_{\rm C}$.
That said, as the cluster ages, $1+t_\nu/t \to 1$ and the effect of the viscous timescale becomes relatively less important, hence the spread in $t_{\rm acc}$ at a given time reduces - this can be seen in the figure since the ends of the solid tracks are closer together than their starts.
\cite{Lodato_2017} also describe how the degree of scatter in $t_{\rm acc}$ at a given time is a proxy for the age of the system relative to the initial viscous timescale and use the observed scatter of the data to constrain the distribution of ages and initial viscous timescales in their population synthesis models.
They note a further consequence of this scatter: for a given initial disc mass at a given time, discs with a large viscous timescale and hence larger $t_{\rm acc}$ have retained larger masses. This leads to an slightly sub-linear relationship between accretion rate and disc mass across a population of discs.

However, we see that for much of the data, which is taken from Lupus, the inferred viscous age (as indicated by the dashed lines) $\frac{M_{\rm disc}}{2\dot{M}_{\rm acc}}<1~\mathrm{Myr}$ i.e. the discs appear too young. A reasonable range of accretion rates is produced, which implies this could be because the models are overmassive.
This discrepancy agrees with the conclusion of \citet{Lodato_2017} that the measured $\langle t_{\rm acc} \rangle \approx 2.5~\mathrm{Myr}$ (which corresponds to an inferred viscous age of $1.25~\mathrm{Myr}$) is too young to agree with the predictions from viscous models if $\gamma<1.2$, given their estimate of the average system age is $\langle t \rangle \approx 1.6~\mathrm{Myr}$. Equivalently, the evolutionary tracks would need a shallower slope $<2.6$ (corresponding to a higher $\gamma$) in order to go through more of the region occupied by the data, rather than passing generally to the right of it at high mass.

\subsection{The Effect of Including Dust Modelling}
The lower left `Dust-to-gas Ratio' panel of Figure \ref{fig:4panels_nope} shows how the relative masses of dust and gas evolve in the model. If the discs were to evolve with a constant dust-to-gas ratio, then the tracks would be parallel to the dashed lines, which represent dust-to-gas ratios that are successive powers of 10. However, it is apparent that although the disc models start with the ISM ratio, due to radial drift the dust masses drop much more rapidly than the gas masses and the discs move towards a lower dust-to-gas ratio. Between 1 and 3 Myr, the dust depletion can be anywhere between a factor of 10 and 1000 for little change in gas mass. This is broadly consistent with the factor $\sim 50$ decline in mass between the median Class 0 dust mass in Perseus and median Class II dust mass in Lupus \citep{Tychoniec_2020}.
From the start/end points of the solid tracks (most easily seen in the blue tracks for $100~\mathrm{au}$ models), we see that at a given time, the dust depletion is more severe for lower disc masses. The dust dynamics are controlled by the Stokes number, which is inversely proportional to surface density (Equation \ref{eq:Stokes}), so in lower mass discs, the grains do not have to grow to so large a size before they have the same dynamics and drift sets in sooner.
Moreover, at a given mass, the more compact a disc, the steeper the tracks and the more severe the dust depletion, as the dust is concentrated close to the star where its radial drift timescale\footnote{For fragmentation limited dust, which is the appropriate limit for small radii at the earliest times \citep{Birnstiel_2012}, this scales as $\tau_{\rm rd} \propto R^{1/2}$.} 
is low.

The lower right `Dust Mass Measurement' panel of Figure \ref{fig:4panels_nope} shows the true dust mass in each model, plotted against the dust mass estimated from the sub-mm flux densities ($ M_{\rm d, obs}$, henceforth "observers' equivalent dust mass") calculated as in Section \ref{sec:obsmasses}.
At 1 and 3 Myr, the solid tracks lie around the dashed line labelled 0.1, indicating that the true dust masses may be around a factor of 10 lower than estimated.
That observed dust masses may be overestimates (except for the most massive or most highly inclined discs) was also the conclusion of the more sophisticated Monte Carlo radiative transfer calculations conducted by \citet{Miotello_2017}.
In our case, there are two factors contributing to this discrepancy.
Firstly, we use an opacity that is also varying in space and time because it depends on the size of the grains, rather than the single value assumed by the estimates (Equation \ref{eq:Mdobs}).
Moreover, the observers' equivalent dust masses are all calculated for a single dust temperature of $20~\mathrm{K}$. For our temperature model (Equation \ref{eq:T_R}), the disc would not reach this sort of temperature until outside $200~\mathrm{au}$, whereas most of the emission would be from dust at smaller radii, and from a range of higher temperatures. This means that less dust is needed to produce the same flux (cf Figure 9 of \citet{Pascucci_2016} where assuming emission at the gas temperature, rather than $20~\mathrm{K}$ produced higher mm fluxes). The effect of temperature can be seen in that the initially large discs are the ones which come closest to an accurate measurement in their evolution, and that between $1$ and $3~\mathrm{Myr}$ the accuracy of the masses gets better as the disc spreads.

This result was already shown for these dust evolution, temperature and opacity models by \citet{Rosotti_2019} who concluded that in practice, when using an opacity appropriate to compact grains, the flux was dominated by material with an opacity that was higher than the value used by observers: the effect of using a larger opacity was more important than any `invisible' mass (either at low opacities or in optically thick regions).
It is worth noting that the opacity is highly dependent on poorly constrained properties such as the composition and porosity of the dust, and this may affect the exact factor to which the observers' equivalent and true masses differ. Similarly the disc temperatures are not well-known, and colder disc models would show less discrepancy between these values (this is explored more in Section \ref{sec:parameters}). However as we shall see, for the observed dust masses to be reconciled with these particular models (without some modification to impede the loss of dust to drift), the disagreement must be, realistically, at least this large. Similarly, \citet{Rosotti_2019} found that a large opacity was necessary to produce fluxes that agreed with the observed flux-radius relationship.

Between the accuracy of the observers' equivalent dust masses, which appear to be overestimates by a factor $\sim 10$, and the decline of the dust-to-gas ratio, which leads to underestimates by a factor of 10-1000 when converting from the dust mass to gas mass, we conclude that the observers' equivalent disc masses may be underestimated by a factor of anywhere between 1 and 100. This effect is largest for the most compact, and least massive, discs. This would mean that the inferred viscous ages are likewise systematically underestimates of the true system age.
Thus we have a mechanism for reducing the previously discussed tensions between the inferred viscous ages predicted by theoretical models and those observed in Lupus.

In the upper right `Observational' panel of Figure \ref{fig:4panels_nope}, we see that the tension is indeed reduced when we compare, like-for-like, the data from Lupus and the models in terms of the observers' equivalent dust mass. The effect of radial drift is to deplete the dust mass more rapidly without affecting the accretion rate, so the model tracks become much flatter and can pass through more of the region occupied by the data. Moreover this happens on appropriate timescales: the solid sections of the tracks (representing the regions of the $\dot{M}_{\rm acc}-M_{\rm dust}$ plane accessible to the models between $1$ and $3~\mathrm{Myr})$ have moved to coincide much better with the main locus of the data. The accessible region is bounded by the location of the most massive discs at high accretion rates, and the least massive at low accretion rates. Moreover, they are bounded at high $t_{\rm acc}$ by the largest discs. The reasons for this are twofold: firstly, the large viscous timescale in these discs, and secondly the relatively inefficient radial drift, which has prevented such a strong depletion in mass.  
Since there is this dependence of drift efficiency on the initial disc radius (viscous timescale), the spread perpendicular to the lines of constant age is much increased compared to the predictions of viscous theory alone. This is consistent with the finding of \citet{Mulders_2017} that a scatter in the dust-to-gas ratio could help explain the scatter in the data, but in this case it occurs in a systematic sense.

Moreover, in the upper right corner, we see that the location in this plane of large massive discs at $1~\mathrm{Myr}$ is little changed from the predictions of viscous theory. However, as we go to increasingly low mass discs we see more of a shift in the location of the solid track sections, consistent with the more efficient drift.
An important consequence of this is that at a given initial disc radius (viscous timescale), the trend of accretion rates with dust masses is flattened slightly below linear, consistent with measurements of the slope in Lupus/Chamaeleon in the range $0.7-0.8$ \citep{Manara_2016,Mulders_2017} - this is reflected in the slope of the upper/lower limits of the accessible regions.

\subsection{Impact of Dust Modelling in Regions with Different Ages}
Between 1 and 3 Myr, the larger discs in the sample continue to decline in dust mass rapidly, whereas in smaller discs the loss of dust has decelerated since the low dust surface densities make the collisional growth timescale of the dust longer and limit the resupply of dust grains large enough to drift.
Thus, the larger discs begin to catch up (as indicated by the blue tracks in the upper-right `Observational' panel of Figure \ref{fig:4panels_nope} being more horizontal than the yellow/orange) - for a coeval population of discs, the spread in $t_{\rm acc}$ decreases as it would for a purely viscous model.

Notably though, the low $t_{\rm acc}$ limit of the region accessible to the models does not evolve significantly because the initially compact discs, which define that limit, also evolve along it. This leads to a constant upper locus at an inferred viscous age of $\sim 0.1~\mathrm{Myr}$.
This key result does not simply arise from the evolution of the true dust masses, for which the limit moves to higher $t_{\rm acc}$ as accretion rates decline, but results from an interplay between this and the relationship between the true and observers' equivalent dust masses, which changes as the dust size and location evolve.

Overall, therefore the size of the accessible region is largely set by the evolution of the largest discs. The length of the solid section of their tracks implies that they spend several Myr in the region of the plot where most of the data are located and should be easily observable there. By contrast, the more extended dashed track to the right of the solid section indicates that the position in this plane evolves more rapidly during the first Myr; thus systems are less likely to be observed in this part of the diagram.

\begin{figure*}
    \centering
	\includegraphics[width=\linewidth]{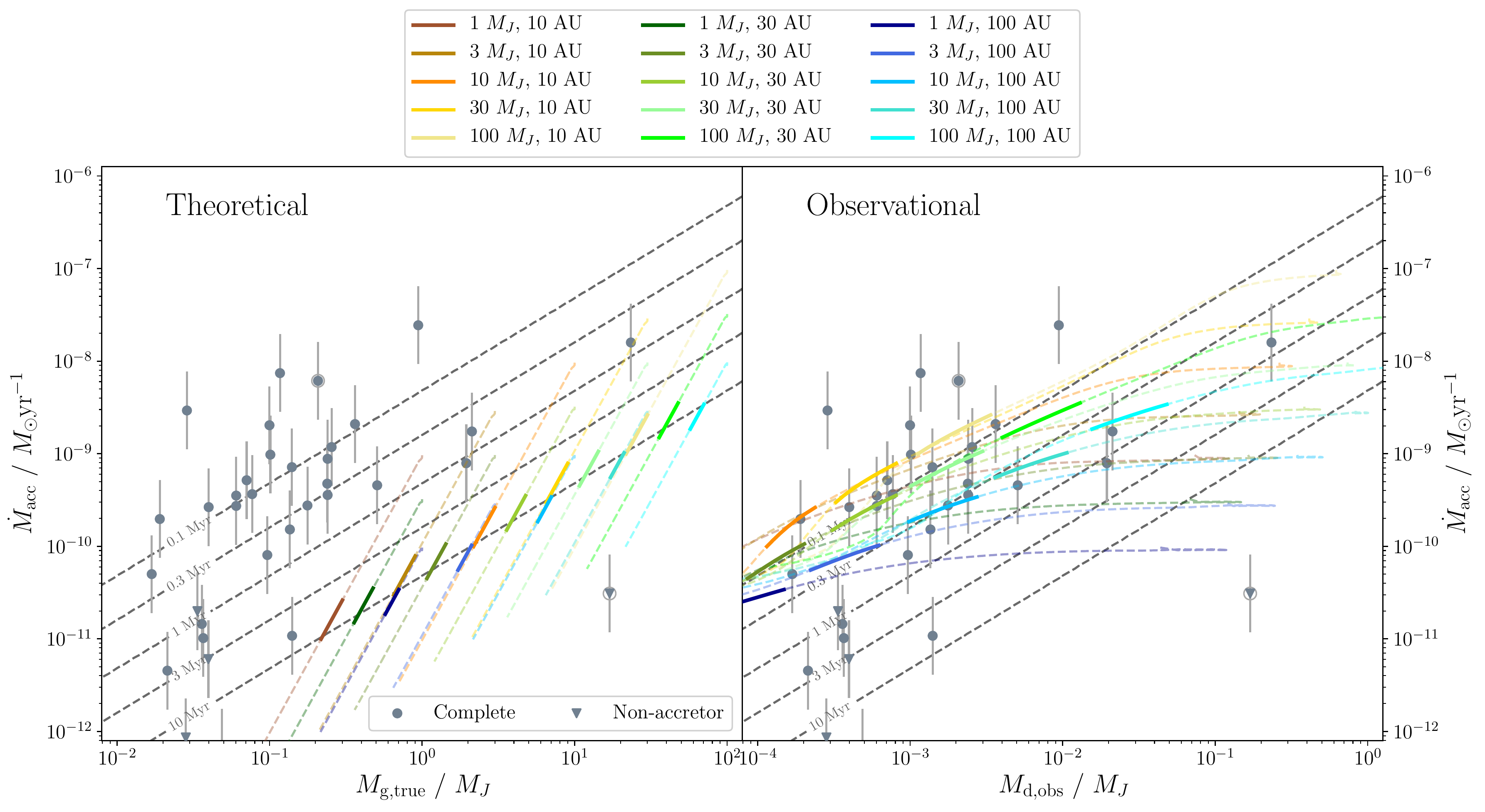}
    \caption{As the upper panels of Figure \ref{fig:4panels_nope} compared to data for Upper Sco from \citet{Manara_2020}, with the dust masses rescaled to use a consistent temperature and opacity across regions. In each panel, the solid section of the tracks indicates the area covered by the models at ages between $5$ and $10~\mathrm{Myr}$.}
    \label{fig:nope_old}
\end{figure*}
We now turn our attention to even older ages, as appropriate to Upper Sco.
Figure \ref{fig:nope_old} compares the same models as discussed so far to the data from \citet{Manara_2020}, with model locations $5$ and $10~\mathrm{Myr}$ highlighted. The data points have had their dust masses rescaled from those quoted by \citet{Manara_2020} (in turn based on \citet{Barenfeld_2016}) such that the temperature and opacity used to infer the masses from the sub-mm fluxes are consistent with those used by \citet{Ansdell_2016} for Lupus, and hence with the method we use to estimate dust masses from our models.

It is clear from the left panel that, as discussed by \cite{Manara_2020}, a purely viscous model would be in stark contrast to the location and spread of the data in the $\dot{M}_{\rm acc}-M_{\rm disc}$ plane if the disc masses inferred using the prescription in \cite{Ansdell_2016} correctly represent the true disc masses. In fact the region of the plane occupied by the models on these timescales lies totally separate to the data.

However, when dust evolution is included in the models and the models are plotted in terms of the observers' equivalent dust masses, as in the right panel, we again see the strong effects of drift which allow a number of the intermediate accretion rate sources to be explained.
Continuing on from what was described between 1 and 3 Myr, the upper locus to the region accessible to the drift models at $t_{\rm acc} \sim 0.1~\mathrm{Myr}$ still hasn't evolved. Thus, the interplay between the effects of radial drift and the changing spatial distribution of the dust provides a good explanation for why in any mass bin, the upper limit to the accretion rates does not seem to change with time \citep{Manara_2020}.
Moreover, the tracks of the larger disc models have turned parallel to this upper locus; the solid portions of the tracks are similarly short to those in Figure \ref{fig:4panels_nope}, indicating that the models spend a large amount of time in the vicinity of the bulk of data.

The range of modelled observers' equivalent dust masses is also able to encompass most of the observed masses. In this context, the larger mass discs seen in Lupus could indeed be understood as the progenitors of the lower mass discs in Upper Sco, with a corresponding decline in the accretion rate of a given source with age. The five sources which lie within the mass range covered by the models, but at too high accretion rates, could be accounted for by external photoevaporation. This process potentially plays a role in disc evolution in Upper Sco \citep{Rosotti_2017,Manara_2020} as there is a considerably higher stellar density than in Lupus \citep{Damiani_2019}, which should result in a number of discs close to high mass O and B stars and therefore exposed to large FUV fluxes \citep{Trapman_2020}.

As it stands, the models also produce some very low-mass discs
at $\sim 10^{-4}~\mathrm{M_{\rm J}}$ with accretion rates just below $10^{-10}~\mathrm{M_{\sun}}~\mathrm{yr^{-1}}$. Since the sample of \citet{Manara_2020} pre-selected only those sources with statistically significant sub-mm detections \citep{Barenfeld_2016}, it is not clear whether these `transparent accretors' are present amongst the sub-mm non-detections in Upper Sco.

\subsection{Summary of the Basic Model}
Comparing the upper left and right panels of Figure \ref{fig:4panels_nope} or the two panels of Figure \ref{fig:nope_old}, we see that by reducing the dust mass through radial drift, the dust evolution model represents a great improvement in reconciling viscous models with observations in both Lupus and Upper Sco. Specifically, it can explain discs that would otherwise appear too young, i.e. have too low a mass given their accretion rate at their likely ages. It also increases the range of possible accretion rates at any one mass to be more consonant with the scatter in $t_{\rm acc}$ seen in the data, and predicts an upper locus at an inferred viscous age of $\sim 0.1~\mathrm{Myr}$ that does not evolve with time.

However there remain a number of data points, particularly those at high inferred viscous ages $\sim 10~\mathrm{Myr}$ in Lupus (i.e. at large masses for their accretion rates), which are not explained by the radial drift models. Moreover, no model with a dust mass high enough to be detected in ALMA surveys is seen to have accretion rates much below $\sim 10^{-10}~\mathrm{M_{\sun}}~\mathrm{yr^{-1}}$, while a number of observed systems in both Lupus and Upper Sco do. In section \ref{sec:parameters} we discuss how lower $\alpha$ values could mitigate these problems.

The discs with the highest $M_{\rm d, obs}$ could be explained by the basic model if they are younger than $1~\mathrm{Myr}$ as radial drift would have had less time to deplete the dust. However, as we have argued, the evolution through this region is very rapid, so we would not expect many sources to be observed in this region. A more reasonable explanation would be that these discs were initially large, with $R_{\rm C}$ greater than the $100~\mathrm{au}$ maximum used so far, such that radial drift becomes even less efficient.

More sophisticated remedies to these conflicts could include: variations on the assumed dust parameters that reduce the speed of radial drift (thus both increasing the dust masses at a given time and extending the time over which discs may be observed at high dust masses) dust-trapping at large radii that ameliorates the depletion due to drift, or the effects of internal photoevaporation. The latter can open a gap at small radii thus trapping remaining dust in the outer disc. Importantly, photoevaporation also acts to reduce the accretion rates, which could help with the inability to produce discs at low accretion rates.
We consider the impact of each of these in the subsequent sections.

\section{Impact of Internal Photoevaporation}
\label{sec:photoevaporation}
In order to avoid a large number of discs with modest accretion rates and low masses we now consider internal photoevaporation, one of the leading candidates for explaining the observed timescales of disc dispersal.

The mass loss rates and profiles due to internal photoevaporation are, however, still quite uncertain \citep{Alexander_2014}, with different models focusing on the influence of irradiation in the X-ray \citep[e.g.][]{Owen_2012,Picogna_2019}, EUV \citep[e.g.][]{Font_2004,Alexander_2006a,Alexander_2007,Wang_2017}, or FUV \citep{Gorti_2009a,Gorti_2009b}. A universal feature of these models is the a{\rm b}ility to open a gap at small radii once the accretion rates fall below the photoevaporation rate, leading to a rapid draining of the inner disc on its viscous timescale with an associated drop in the accretion rate \citep[e.g. the `ultraviolet switch' of][]{Clarke_2001}. Moreover, X-rays appear to be capable of causing a preceding period of `photoevaporation-starved accretion' \cite{Drake_2009}. The draining of dust from the inner disc could lead to the deficit in near-IR excess that characterises transition discs.

Moreover, since most of the data points that we do not explain with our fiducial drift model lie at low accretion rates for a given mass, and several are transition discs, it is therefore natural to ask whether internal photoevaporation could move the models towards explaining the data. Thus we explore the effects of two photoevaporation models - the EUV model from \cite{Alexander_2007} and the X-ray model of \cite{Picogna_2019}. In both cases we assume no dust is removed by the wind.

\subsection{EUV Photoevaporation}
\label{sec:euv}
\begin{figure*}
    \centering
	\includegraphics[width=\linewidth]{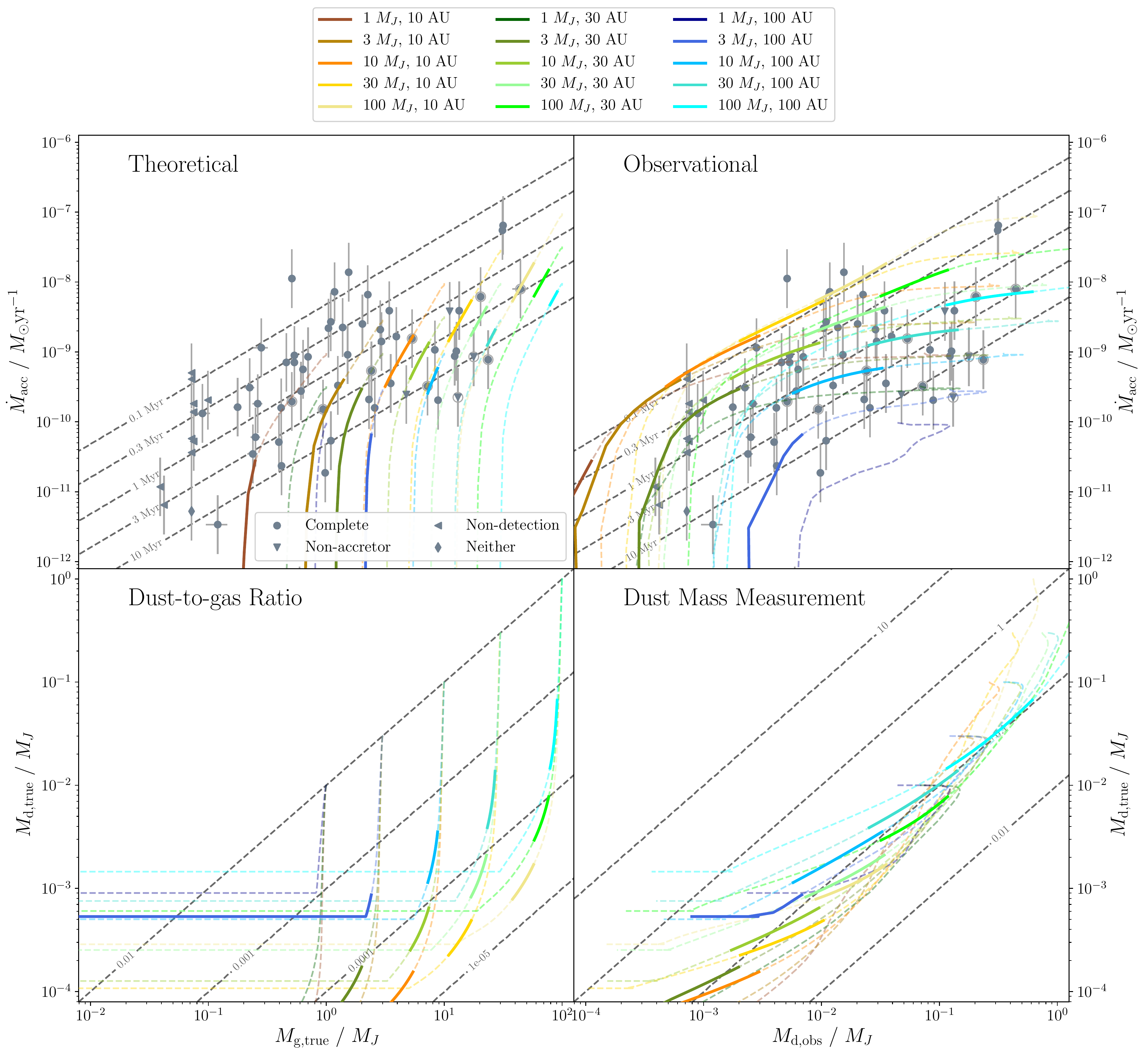}
    \caption{As with Figure \ref{fig:4panels_nope}, but with EUV photoevaporation produced by an ionising photon flux $\Phi = 10^{42}~\mathrm{s}^{-1}$. The solid sections indicate system ages of $1-3~\mathrm{Myr}$.}
    \label{fig:4panels_euv}
\end{figure*}

The EUV photoevaporation prescription given in \citet{Alexander_2007} is a function of the gravitational radius $R_g = GM_*/c_{\rm S}^2$ and the stellar ionising flux $\Phi$, for which a typical value is $10^{42}~\mathrm{photons~s^{-1}}$ \citep{Pascucci_2009}. For a solar mass star, this corresponds to a total mass loss rate of $4.05 \times 10^{-10}~\mathrm{M_{\sun}}~\mathrm{yr^{-1}}$. We ran models for the same parameters as before, and present our results in Figure \ref{fig:4panels_euv}.

In the upper left panel, the evolutionary tracks are no longer straight lines, but curve downwards. This is indicative of the accretion rate being starved, then switched off, by the photoevaporative wind. In the lower left panel, we see that the dust mass eventually levels off - this is due to remnant dust being trapped outside of the gap that is opened at small radii. Once the inner disc drains, the gas in the outer disc continues to be cleared by the direct field \citep{Alexander_2006a,Alexander_2006b}, thus increasing the radius of the inner hole. Consequently, the pressure maximum in which the dust is trapped moves outwards, sweeping the dust out to large radii where the approximation $T=20~\mathrm{K}$ becomes better and the mass determinations become more accurate.

In the upper right panel, we see that the photoevaporating models do represent a further improvement in terms of which observed discs could be explained.
A small number of the discs on the right of the plot, with masses previously too large for their accretion rate, are now accessible to the models between $1$ and $3~\mathrm{Myr}$ since the accretion rates are reduced, though many remain unexplained.

More importantly, the evolution is no longer towards very low masses and moderate accretion rates at late times. Instead, at dust masses $M_{\rm dust}<10^{-2}~\mathrm{M_{\rm J}}$, the tracks extend the accretion rates to previously inaccessible values (lower than $\sim 10^{-10}~\mathrm{M_{\sun}}~\mathrm{yr^{-1}}$).
This allows the models to evolve through the region occupied by the data points at low accretion rates: at first glance the best agreement with the indicated masses would generally seem to be for the $30~\mathrm{au}$ discs or the more massive $100~\mathrm{au}$ discs.
Although the tracks for the $10~\mathrm{au}$ discs extend into regions that have lower masses than the datapoints shown, it is worth noting that there are a number of sources with upper limits on the dust mass (with the markers placed at the upper allowed limit) and these could turn out to be compatible with such evolutionary tracks.
Thus, the data do not currently rule out these sources being in agreement with the most compact disc models, although high sensitivity dust measurements would be required to confirm this.

However, the question of timing is important.
It can be seen that four of the models pass through appropriately low accretion rates $<10^{-10}~\mathrm{M_{\sun}}~\mathrm{yr^{-1}}$ on the $1-3~\mathrm{Myr}$ timescale. These are the lowest mass discs, which are preferentially removed by photoevaporation as found by \citet{Somigliana_2020}.
While we don't necessarily capture the exact masses to reproduce the data, this is, in part, a product of the coarse sampling of the disc parameters: some combinations of masses and radii intermediate to those depicted, or other EUV fluxes, would pass through the right region at just the right time.
The solid tracks here are much longer than in the non-photoevaporating models, implying a more rapid evolution. Thus how likely we are to observe discs at low accretion rates depends on how rapidly the discs evolve through the region relative to their age - it typically takes a few tenths of a Myr for the accretion rates to decline from $10^{-10}$ to $<10^{-12}~\mathrm{M_{\sun}}~\mathrm{yr^{-1}}$, so it is certainly not inconceivable that 10-20 per cent of the discs should lie at such low accretion rates Moreover, $\sim18$ per cent of the Lupus data lie below the track for the non-photoevaporating model with $1~\mathrm{M_{\rm J}}$ and $100~\mathrm{au}$ compared to $\sim29$ per cent of the discs surviving to at least $1~\mathrm{Myr}$ passing through this region on the right timescale, meaning lower chances of observing discs at low accretion rates are not irreconcilable.

Thus, it would seem that EUV photoevaporation can do a good job of producing the low accretion rate discs in the right mass range and avoiding arbitrarily low mass discs that are still accreting. Performing a full population synthesis, with a more realistic, continuous, distribution of masses, radii and observation times - rather than a parameter space investigation - would be necessary to more precisely quantify how well observations of the lowest accretion rates are reproduced by these models. For example, a distribution more strongly weighted towards initially low mass discs would further ameliorate any concerns about whether discs spend enough time at low accretion rates to be observed.

We also note that though the lowest mass large disc seems to cross a region at relatively high mass for low accretion rates, where no discs are observed, it does so on timescales shorter than $1~\mathrm{Myr}$.
Thus, such discs cannot be conclusively ruled out as existing in the initial population.

\subsection{X-ray Photoevaporation}
\label{sec:xray}
\begin{figure*}
    \centering
	\includegraphics[width=\linewidth]{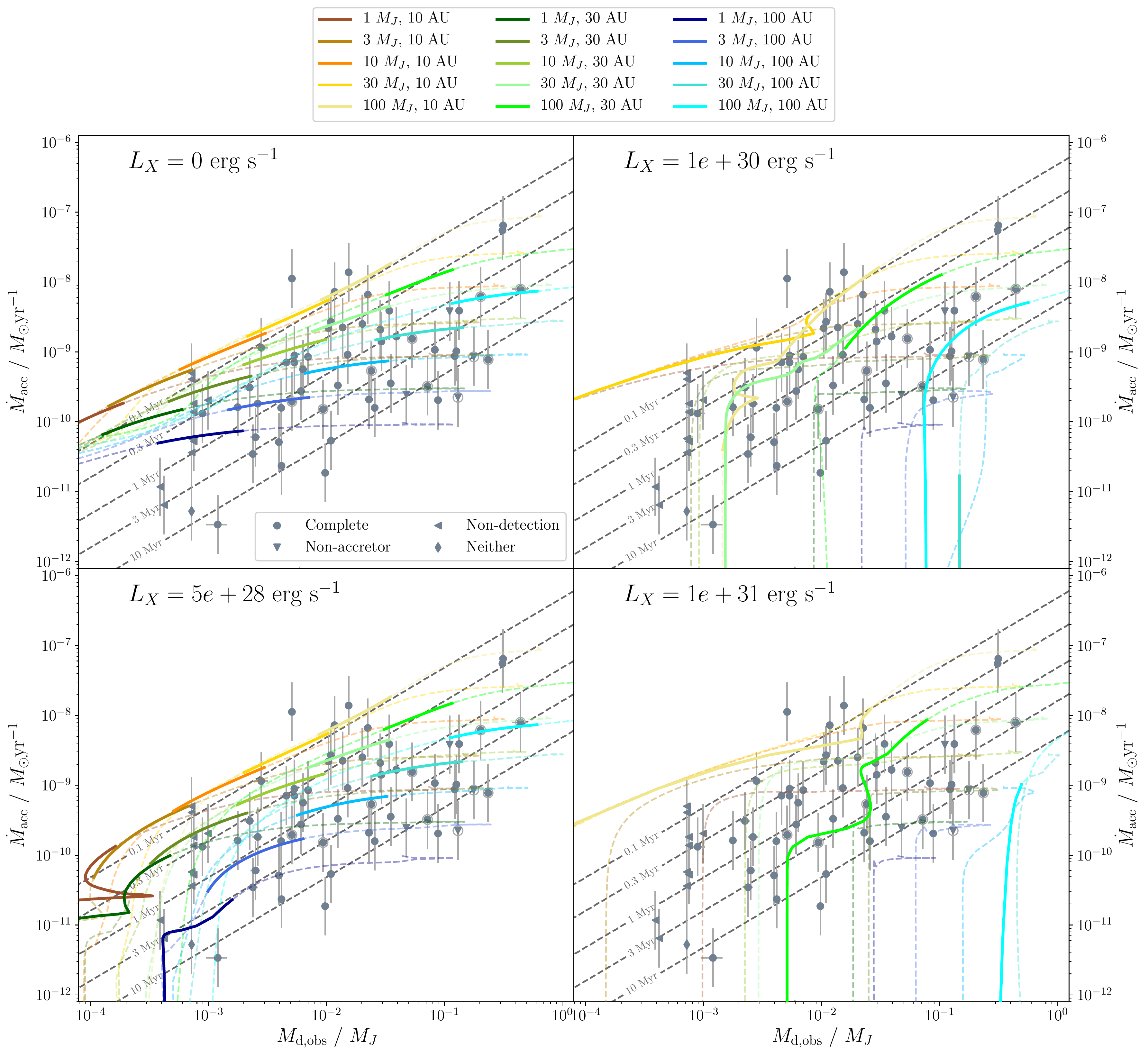}
    \caption{As with the upper right panel of Figure \ref{fig:4panels_nope}, but with X-ray photoevaporation produced by different X-ray luminosities.}
    \label{fig:4panels_LX}
\end{figure*}

The prescription for the X-ray photoevaporation rates given in \cite{Picogna_2019} has a mass-loss rate determined by the X-ray luminosity of the star (Equation \ref{eq:Picogna_Mdot}). Observationally, X-ray luminosities lie roughly in the range $5\times10^{28} < L_{\rm X} / \mathrm{erg~s^{-1}} < 10^{31}$, with a median value of $1-2 \times 10^{30}~\mathrm{erg~s^{-1}}$, and are correlated with stellar mass \citep{Preibisch_2005,Gudel_2007}. In order to span this range, we test luminosities of $5\times10^{28}~\mathrm{erg~s^{-1}}$, $10^{30}~\mathrm{erg~s^{-1}}$, and $10^{31}~\mathrm{erg~s^{-1}}$. We note that the low and high values are close to the mean values for stars of mass $0.1~\mathrm{M_{\sun}}$, $3~\mathrm{M_{\sun}}$ respectively \citep{Gudel_2007}, roughly the stellar mass span of the Lupus data sample.
The evolutionary tracks under each X-ray luminosity are compared in Figure \ref{fig:4panels_LX}.

The models subjected to $L_{\rm X} = 5\times10^{28}~\mathrm{erg~s}^{-1}$ are presented in the lower left panel of Figure \ref{fig:4panels_LX}; this luminosity produces a mass loss rate of $3.71\times10^{-10}~\mathrm{M_{\sun}}~\mathrm{yr^{-1}}$, similar to that in the EUV model presented above.
The evolution is qualitatively very similar to the EUV case.
The mass loss due to EUV photoevaporation is concentrated into a narrower range of radii than in the X-ray case. Consequently X-rays are slightly slower at opening a gap in the disc.
This means that more dust can deplete before accretion is switched off, which is most obvious in that the initially large discs are dispersed at slightly lower masses, perhaps moderately more consistent with the data. Nevertheless, there is little difference between these models and those using an EUV prescription.

For $L_{\rm X} = 10^{30}~\mathrm{erg~s}^{-1}$ (upper right panel of Figure \ref{fig:4panels_LX}), the area spanned by the evolutionary tracks largely fails to reproduce the data.
Very few discs have sufficiently efficient radial drift for the dust depletion to be consistent with the sources which only have upper limits on the mass. Those that do are a very limited range of low mass, compact, discs, which have low enough densities but large enough mass-loss rates in the outer disc that the X-ray driven photoevaporation clears them from the outside in, rather than by opening a gap. Consequently, rather than trapping dust, the lowering of gas surface densities raises the Stokes number of the dust and accelerates the loss of dust to drift; the discs disperse at much lower masses ($M_{\rm d,obs}\sim 10^{-6}-10^{-5}~\mathrm{M_{\rm J}}$) than the upper limits.
However, this paucity of discs reproducing the sources without sub-mm detections may not be a large issue, as lower mass discs tend to be found around lower mass stars, for which the lower luminosities should be more typical.

Another potential issue is that the initially large discs are dispersed at relatively high dust masses and all pass through a region not occupied by the data. As with the EUV models that did similarly, this is not problematic if they do so on timescales shorter than $1~\mathrm{Myr}$; however the more massive of these discs do survive more than $1~\mathrm{Myr}$.
It is not, however, impossible that some discs do reside in this area. Some sources (plotted as downward pointing triangles) in the right mass range are classed as potentially non-accreting - as they have accretion luminosity consistent with chromospheric noise \citep{Alcala_2017,Manara_2020}) - and hence their measurements could be considered upper limits. However at least one of these, MY Lup, appears to be more strongly accreting when line luminosities rather than continuum luminosities are used \citep{Alcala_2019}. 
It is likewise possible that some of the discs that we consider too massive to agree with the basic dust evolution models could also be explained as initially large discs shortly before their dispersal by photoevaporation.

Finally, we consider $L_{\rm X} = 10^{31}~\mathrm{erg~s}^{-1}$ (lower right panel of Figure \ref{fig:4panels_LX}), which is really most appropriate for a relatively massive $\sim3$ solar mass star - a rare occurrence at the upper end of the observed range. In this case, only the most massive discs survive to 1 Myr.
Considering for a moment a hypothetical scenario in which these high mass loss rates are appropriate for all stellar masses, then to resolve this timescale issue, we might invoke the cluster being younger than measured. Then, however, as with $L_{\rm X} = 10^{30}~\mathrm{erg~s}^{-1}$, it would once again be very hard to explain the lowest mass discs: with this mass loss rate, the gap usually opens before radial drift has long enough to deplete the observers' equivalent dust mass to $\lesssim 10^{-2} \mathrm{M_{\rm J}}$. Such a high mass loss rate across the cluster is thus incompatible with both the timescales and masses of the discs in the region, which given the uncertainties in the photoevaporation models, puts a useful constraint on the acceptable mass loss rates.
Conversely, if these mass loss prescriptions are right, then the massive stars to which they are thought to apply must start with very massive discs, or be very young to avoid being dispersed on timescales shorter than the age of the Lupus region.

\subsection{The Role of Internal Photoevaporation in Upper Sco}
\begin{figure*}
    \centering
	\includegraphics[width=\linewidth]{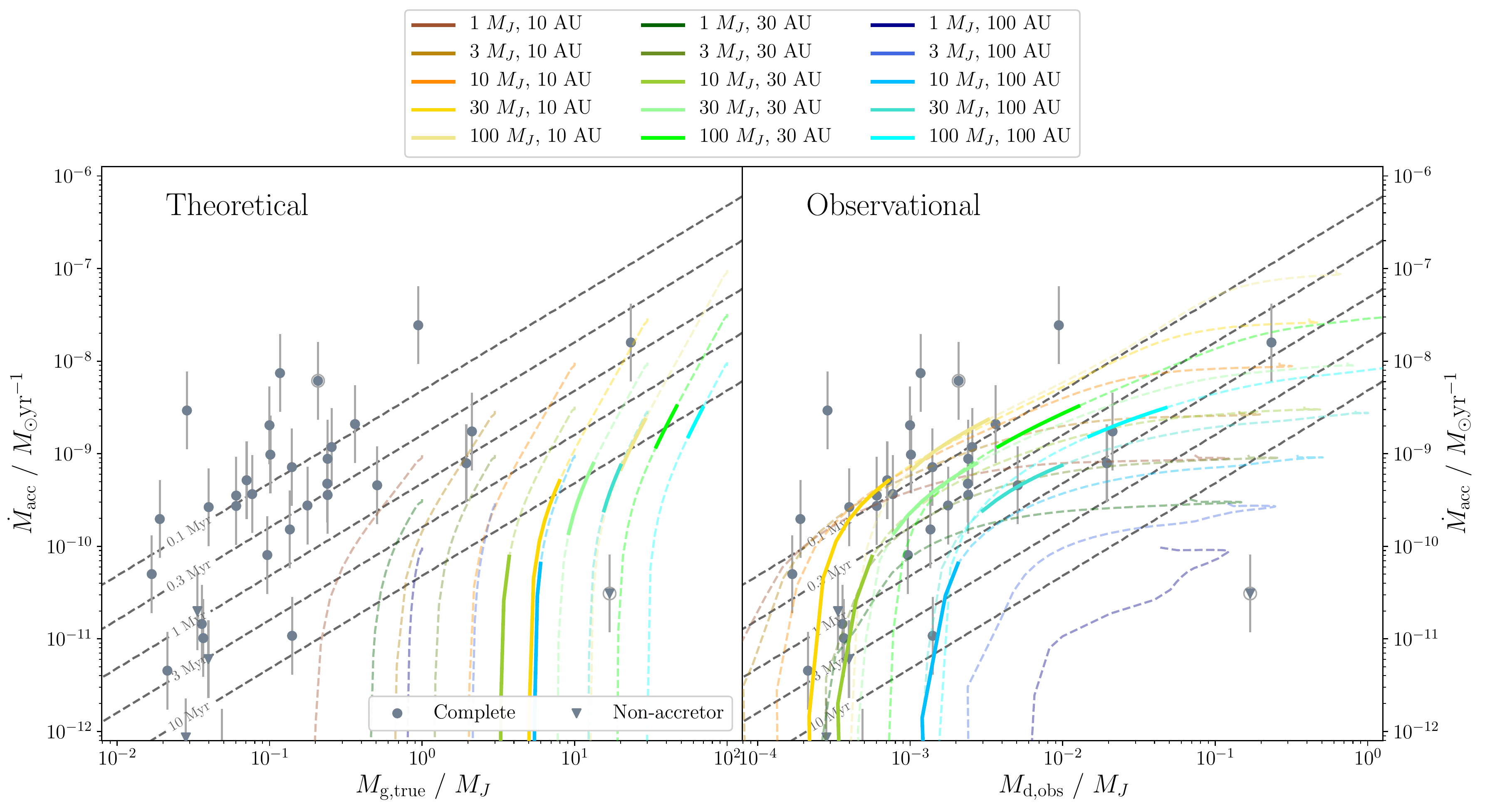}
    \caption{Models with EUV photoevaporation produced by an ionising photon flux $\Phi = 10^{42}~\mathrm{s}^{-1}$. As the upper panels of Figure \ref{fig:4panels_euv} but with data from \citet{Manara_2020}. The solid sections indicate system ages of $5-10~\mathrm{Myr}$.}
    \label{fig:euv_old}
\end{figure*}
Figure \ref{fig:euv_old} shows the EUV photoevaporation models from Figure \ref{fig:4panels_euv} at 5 and 10 Myr, compared to the discs in Upper Sco \citep{Manara_2020}.
As was the case at earlier times, the effects of photoevaporation are to extend the area accessible to the models down to lower accretion rates, and to prevent discs reaching very low masses, which are not present in the current sample.
We reiterate that since only cluster members with sub-mm disc detections were surveyed by \cite{Manara_2020}, at the lowest masses we lack information about how the plane is populated, specifically whether the accretion rates are high (`transparent accretors'), or whether they are more consistent with being starved by photoevaporation - this information would be useful for constraining the models.

Since photoevaporation becomes effective once the accretion rate becomes of order the mass loss rate, a given photoevaporation rate sets a floor scale to the accretion rates, beyond which the decline in the accretion rate and subsequent dispersal of the gas disc happens rapidly.
Thus, if we assume that at any time at least some discs should have accretion rates approaching the photoevaporation rate and thus be approaching dispersal, we would expect not to see an evolution in the lower limit of the observed accretion rates.
Note that a similar suggestion was made by \citet{Ercolano_2014}: that the trend between accretion rates and stellar mass could reflect the dependence of the X-ray photoevaporation rate on stellar mass (principally via the X-ray luminosity) if one assumes that we are most likely to see the discs just before the photoevaporative wind opens a gap and hence disc accretion rates are most likely to be recorded at their lowest value before dispersal.

This gives a very satisfying explanation for the similar distributions of accretion rate in any one mass bin between Upper Sco and younger regions. The upper limit is set by the non-evolving upper locus at $\sim 0.1~\mathrm{Myr}$ produced by the radial drift dust model as discussed in Section 3, while the lower limit is set by the onset of rapid dispersal by photoevaporation.

\section{Impact of Parameter Choices}
\label{sec:parameters}
\subsection{Dust Growth Model Parameters}
\citet{Appelgren_2020} also recently conducted a study into the evolution of the gas and dust components throughout the lifetime of a disc, however using dust of a fixed $100 ~ \mathrm{\mu m}$ size. They found that three phases - a short formation phase in which the gas and dust masses of the disc are increased, a slow-dust-drift phase in which the dust is well-coupled to the gas and a final rapid drift phase in which the dust disc clears - could agree well with observed dust disc masses in young clusters. This last phase sets in once the Stokes number of the dust grows large enough for it to dynamically decouple from the gas, which for a fixed size requires the gas surface density to be sufficiently low (see Equation \ref{eq:Stokes}). Crucially, they found this last phase occurred for a very narrow range of accretion rates, which runs somewhat contrary to our findings that drift allows for a large spread in $t_{\rm acc}$ and therefore at a given mass, the accretion rate. We interpret this as being due to the fact that in the similarity solutions the time at which the surface density becomes low enough in enough of the disc for drift to be significant is longer for initially high masses and compact discs, giving the accretion rate more time to fall. On the other hand, the accretion rates start higher for these discs - the two effects cancel somewhat to give a weak dependence on the disc's initial state. Conversely, since in our model the dust can grow much larger, it starts drifting after only a short growth phase, and thus at an accretion rate determined largely by the initial conditions. Note that as dust is depleted, the maximum grain size declines again until it hits the minimum size we impose. After a period of inefficient drift (where the curves flatten out in the lower left panel of Figure \ref{fig:4panels_nope}), the surface densities become low enough that drift accelerates and the dust behaves like the fixed size dust of \cite{Appelgren_2020}; this is why all of our models leave the plot at very similar accretion rates around $\sim10^{-10}~\mathrm{M_{\sun}}~\mathrm{yr^{-1}}$.

\begin{figure}
    \centering
    \includegraphics[width=\linewidth]{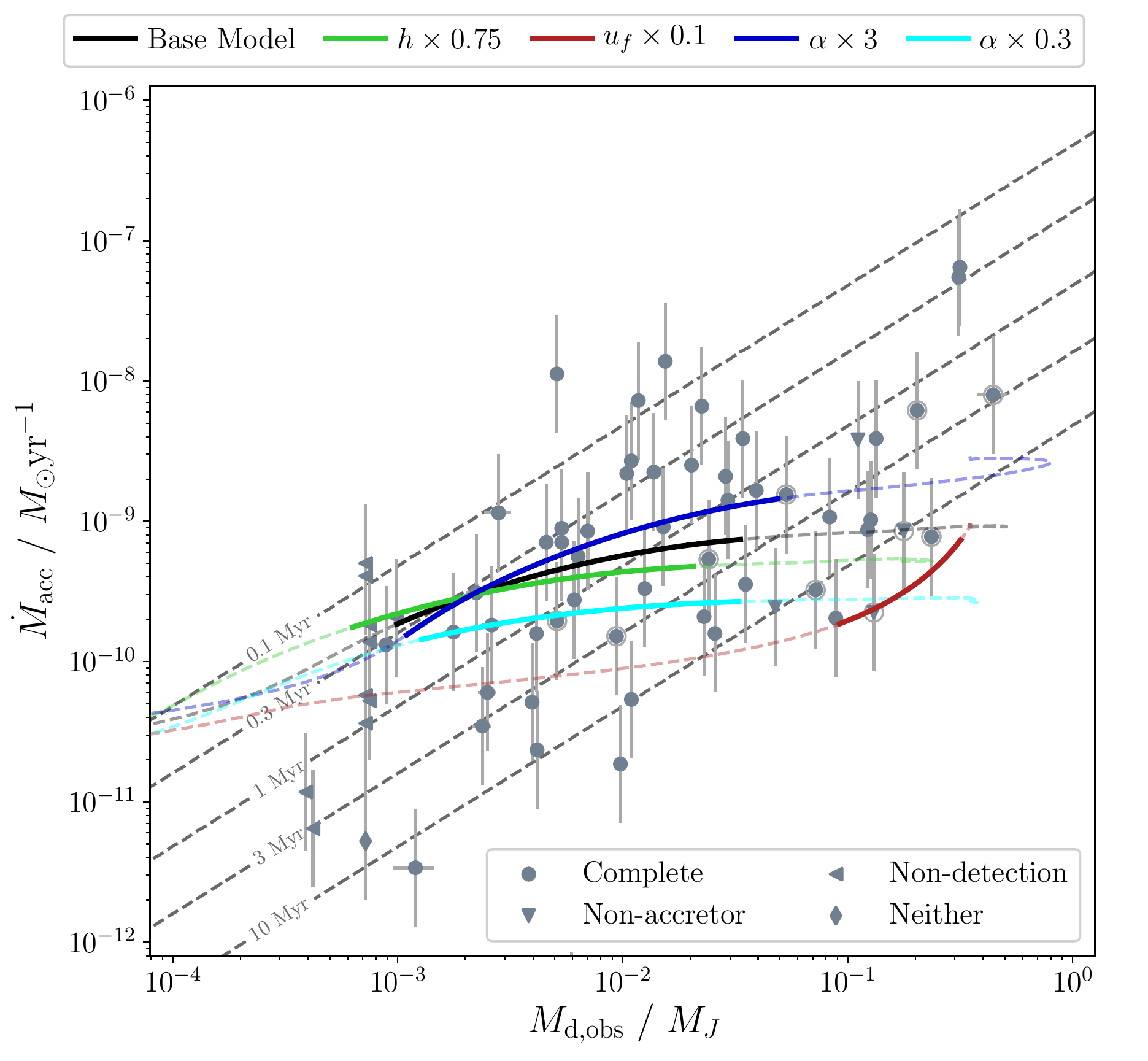}
    \caption{Varying the parameters of the dust model for a $100~\mathrm{au}$, $10~\mathrm{M_{\rm J}}$ disc. The solid sections indicate system ages of $1-10~\mathrm{Myr}$.}
    \label{fig:vary_dust}
\end{figure}

The dust evolution models we present depend on a number of parameters, in particular those listed in Table \ref{tab:dust_model}. Changing these parameters will have an effect on the size to which dust can grow, and its degree of coupling to the gas (affecting the efficiency of radial drift) so it is important to consider the effects of our choices. Figure \ref{fig:vary_dust} shows the results of models with changed parameters for an initial large ($R_{\rm C}=100~\mathrm{au}$) and medium mass ($M_{\rm disc,0}=10~\mathrm{M_{\rm J}}$) disc. The parameters have generally been changed in such a way as to attempt to reduce the efficiency of radial drift, in order to see if this can account for the apparently old discs at large masses that are otherwise unexplained by these dust models.

\begin{figure*}
\begin{subfigure}{0.49\linewidth}
    \centering
    \includegraphics[width=\linewidth]{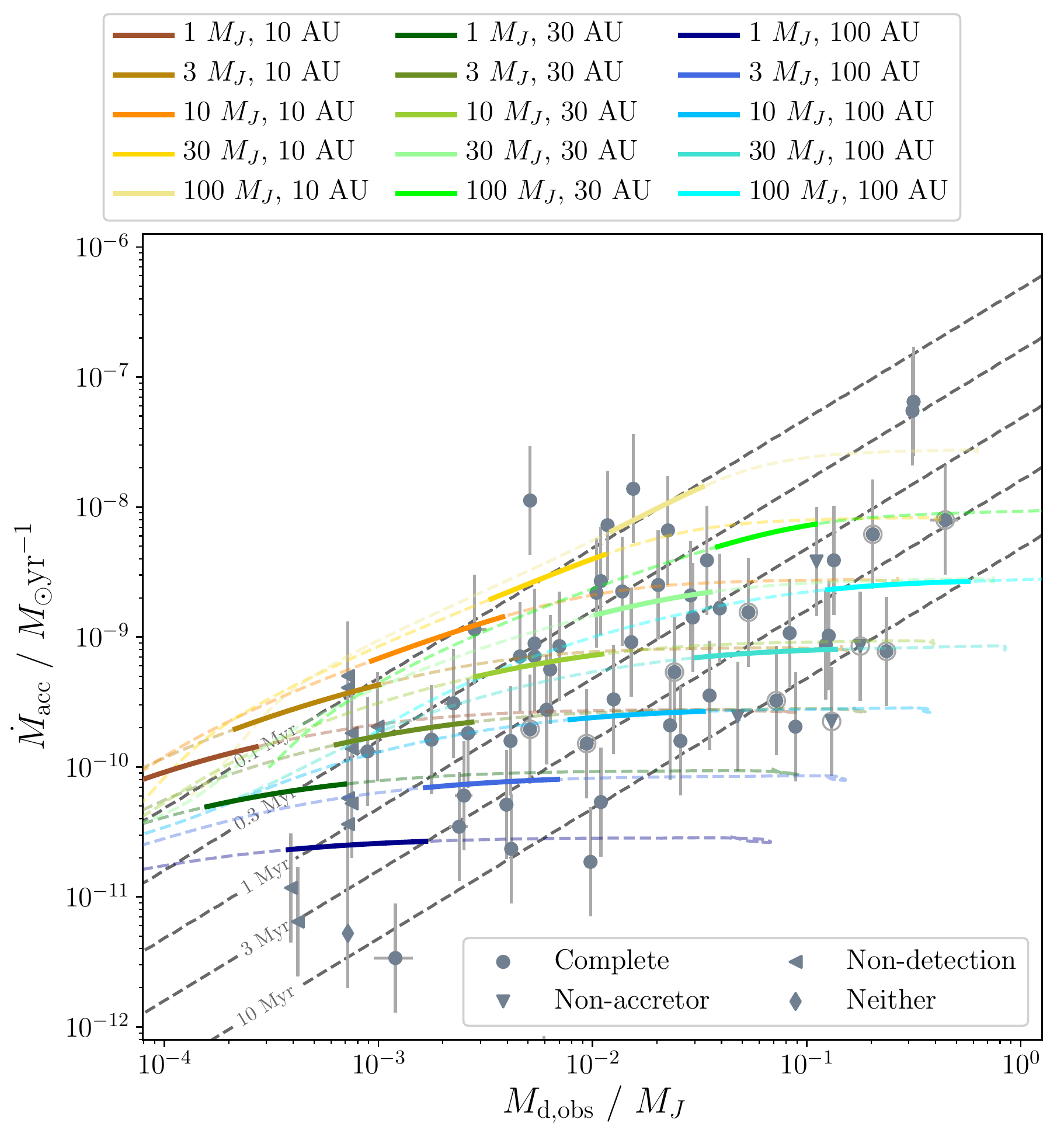}
    \caption{As with the upper right panel of Figure \ref{fig:4panels_nope} but with $\alpha=3\times10^{-4}$.}
\end{subfigure}
\begin{subfigure}{0.49\linewidth}
    \centering
    \includegraphics[width=\linewidth]{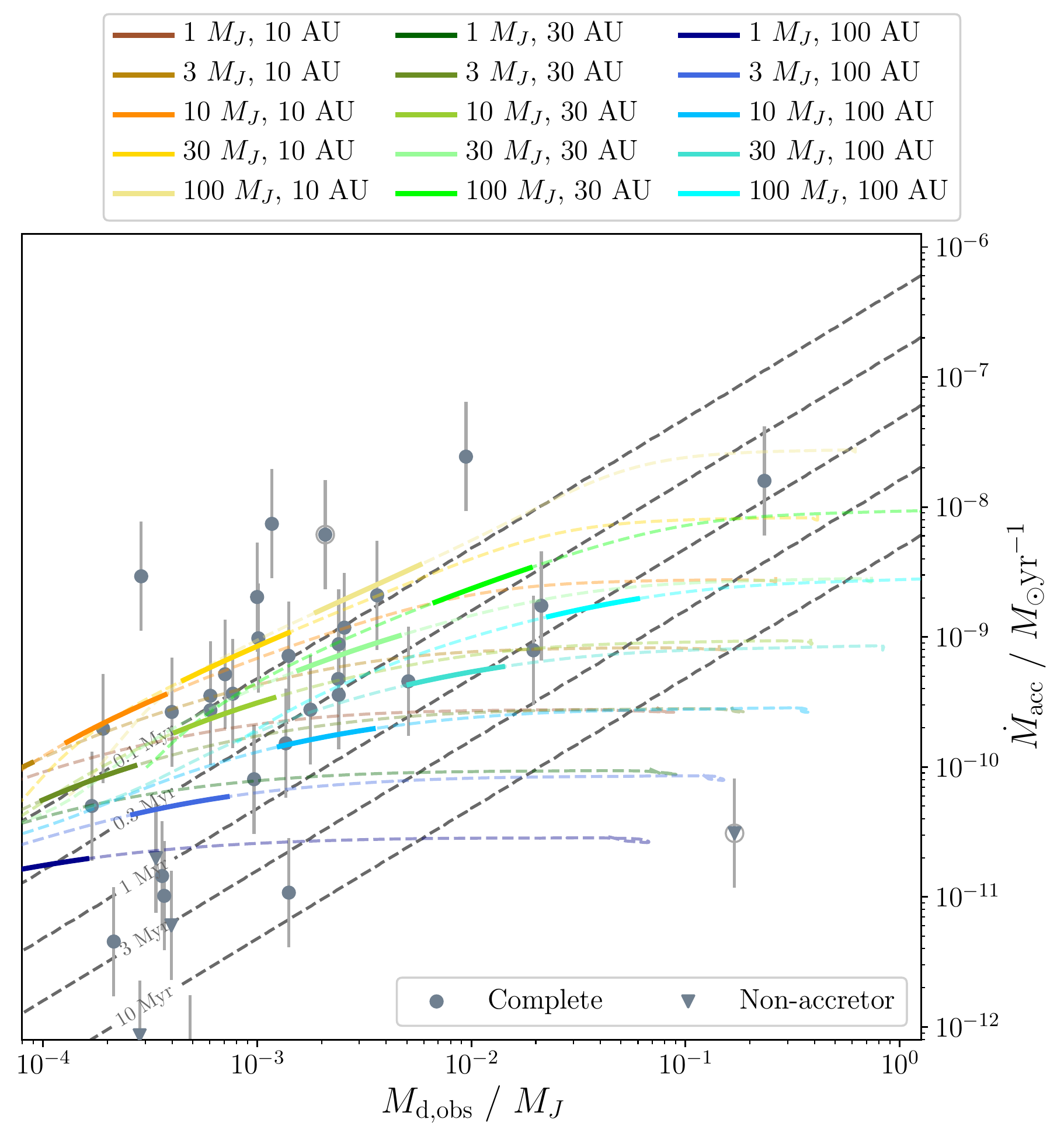}
    \caption{As with the right hand panel of Figure \ref{fig:nope_old} but with $\alpha=3\times10^{-4}$.}
\end{subfigure}
\caption{}
\label{fig:vary_alpha}
\end{figure*}

\begin{figure*}
\begin{subfigure}{0.49\linewidth}
    \centering
    \includegraphics[width=\linewidth]{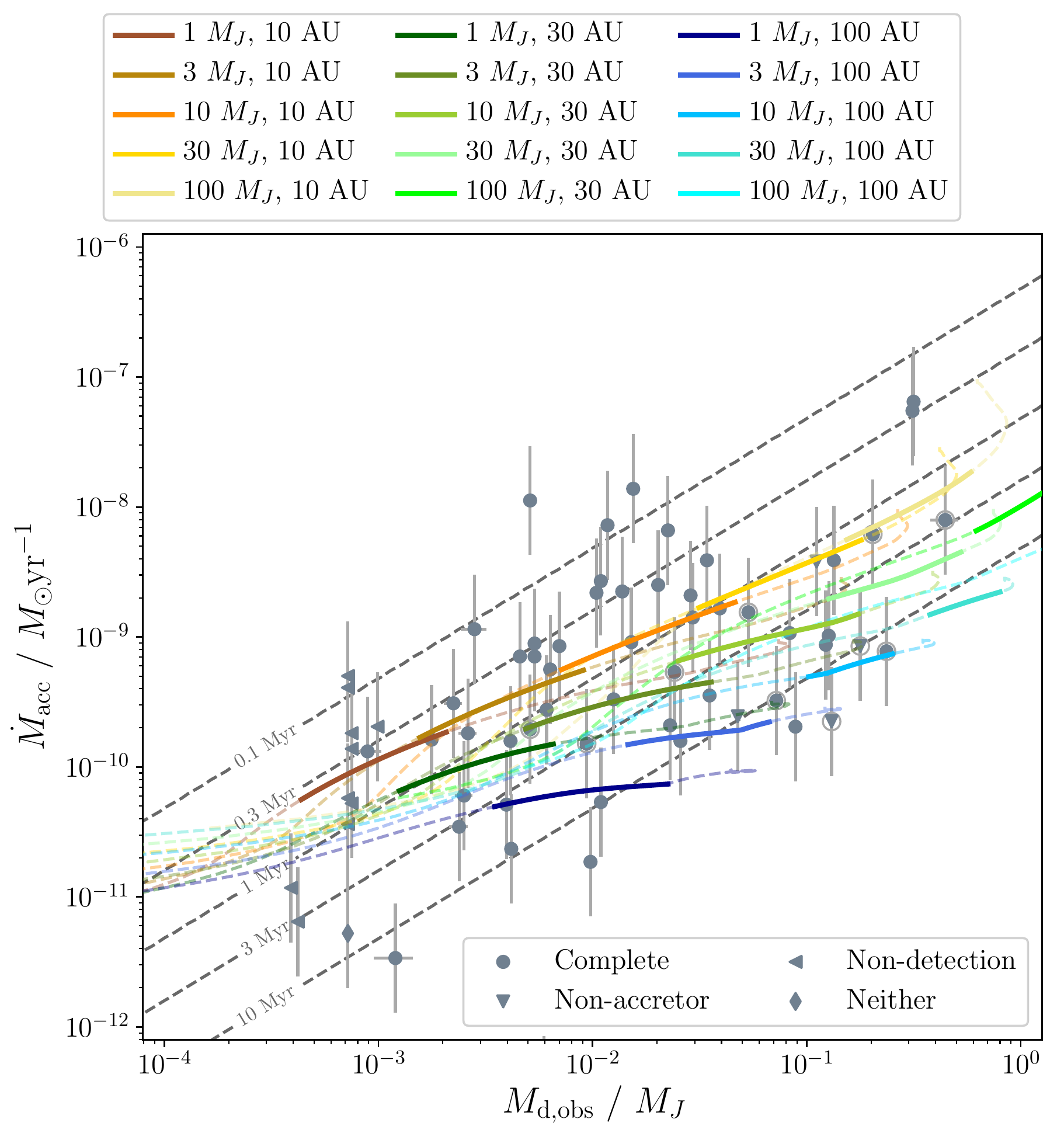}
    \caption{As with the upper right panel of Figure \ref{fig:4panels_nope} but with $f_{\rm grow}=10$.}
\end{subfigure}
\begin{subfigure}{0.49\linewidth}
    \centering
    \includegraphics[width=\linewidth]{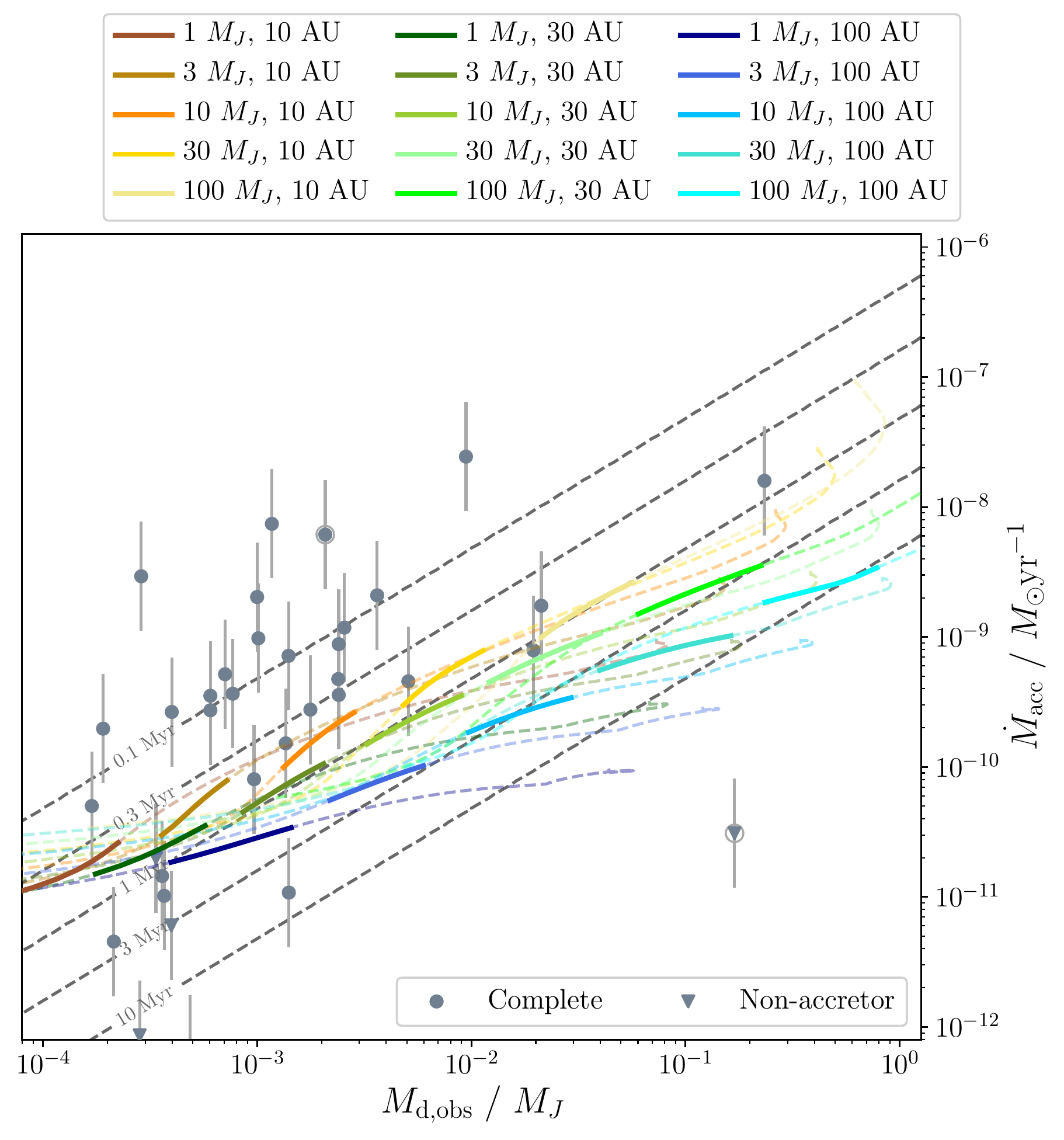}
    \caption{As with the right hand panel of Figure \ref{fig:nope_old} but with $f_{\rm grow}=10$.}
\end{subfigure}
\caption{}
\label{fig:vary_fgrow}
\end{figure*}

The fragmentation limit of $10~\mathrm{m~s^{-1}}$ that we adopted is appropriate for dust grains with an icy composition \citep{Gundlach_2015}. However, grain composition should be expected to vary throughout the disc as the chemistry changes, particularly across snowlines. For less icy grains, the grains should be weaker and more easily fragmented, with fragmentation velocities more like $1~\mathrm{m~s^{-1}}$ \citep{Blum_2008}.
A model with this lower value throughout the disc is presented as the dark red track in Figure \ref{fig:vary_dust}.
Since the Stokes number (size) for fragmentation-limited dust scales as $u_f^2$ (Equation \ref{eq:afrag}), then non-icy grains should be better coupled to the gas. Indeed, we see that this model leads to initially much slower evolution of the dust mass, as shown by the initially very steep track (in contrast with the shallow tracks in Figure \ref{fig:4panels_nope}), until such time as the gas surface densities become low enough for drift to be significant. Thus having some discs with more easily fragmented dust could explain the discs observed to have higher masses than the drift models. However, without higher $u_f$, the difference in the efficiency of radial drift with disc radius would likely not be strong enough and so on the timescales of Lupus the overall population would look too similar to the purely viscous models.

Since a remaining cause of uncertainty in disc evolution is the origin and magnitude of the angular-momentum transport, usually assumed to be viscosity \citep[e.g.][]{Mulders_2017,Rafikov_2017}, it is worth considering the impact of choosing a different $\alpha$. Moreover, a number of measurements of $\alpha$ have suggested the values could be lower than we have assumed so far \citep{Teague_2016,Flaherty_2017,Fedele_2018,Flaherty_2020}.
A higher $\alpha$ will raise accretion rates, but also make the dust collision velocities larger, leading to smaller, better-coupled, dust grains which could keep the dust mass higher. Conversely, a lower $\alpha$ will lower the accretion rate but could lead to worse-coupled dust.

To see which effects dominate, we ran models with both $\alpha\times3$ and $\alpha\times0.3$, which are represented by the dark and light blue lines respectively in Figure \ref{fig:vary_dust}. It turns out that an increase of a factor of 3 is not sufficient to hold onto enough dust, hence the model has a lower $t_{\rm acc}$ due to the higher accretion rates at $1~\mathrm{Myr}$ and cannot help explain the data points at high $t_{\rm acc}$. On the other hand, the lower $\alpha$ does help extend the range that the models can reach to lower accretion rates.
This is because although the lower $\alpha$ allows inner disc dust to grow to a larger, more rapidly drifting fragmentation-limited size, this phase lasts only as long as it can be resupplied from the outer disc where the dust evolution is drift limited and thus independent of $\alpha$. Thus on timescales of $\sim1~\mathrm{Myr}$ the dust retention is only very weakly dependent on $\alpha$.
To confirm the effect of this change on the whole population we run the whole range of models with $\alpha=3\times10^{-4}$ and show the results in Figure \ref{fig:vary_alpha}.
For the more compact discs, most of the dust is at fragmentation-dominated radii and can drift more effectively resulting in little effect on $t_{\rm acc}$. Across the family of models this thus means that at lower viscosities, we can achieve a larger spread in accretion rates and cover more of the $\dot{M}_{\rm acc}-M_{\rm dust}$ plane\footnote{Conversely a sufficiently higher viscosity would result in all the dust being well-coupled to the gas and eliminate the differential efficiency of radial drift.}. Thus we conclude that viscosities $\alpha\lesssim10^{-3}$ provide a suitable match to the data.

The final parameter we consider is the fraction of collisions leading to grain growth, represented by its inverse $f_{\rm grow}$. This parameter has been used in models of pebble accretion \citep{Ormel_2017}. Moreover \citet{Booth_2020} explored whether $f_{\rm grow} > 1$ could rectify the problem that dust evolution models frequently deplete dust on a faster timescale than observations suggest, specifically that for ]$f_{\rm grow}=1$ dust depletes before gap opening by giant planets could create a transition disc.
For $f_{\rm grow}=10$, they found that dust was removed more slowly and this could help explain the depletion of refractory elements in the Sun.
This slower removal arises because $f_{\rm grow}>1$ slows the growth of dust, which for drift limited dust, also increases its drift timescale by a factor of $f_{\rm grow}$.
Hence, for $f_{\rm grow}=10$ we find that more dust is retained, enough to explain some of the most massive discs with inferred viscous ages around $10~\mathrm{Myr}$.
However, a similar impediment to drift is expected regardless of initial disc properties; comparing Figure \ref{fig:vary_fgrow}a to the upper right panel of Figure \ref{fig:4panels_nope} shows that the dust mass is always increased by a factor $\sim10$ at 1 Myr by increasing $f_{\rm grow}$ to $10$. This can be understood in that the radial drift velocity $v_{rd} \sim \epsilon v_{\rm K} / f_{\rm grow}$, where $\epsilon$ is the local dust-to-gas ratio so $\dot{M}_{\rm dust} \propto \Sigma_{\rm dust}^2 / f_{\rm grow} \propto M_{\rm dust}^2 / f_{\rm grow}$. For $M_{\rm dust} \ll M_{\rm dust,0}$ (and assuming the gas densities change negligibly over this period), this integrates to $M_{\rm dust} \propto f_{\rm grow}/t$. The effects of mass or radius on these simple assumptions would seem to be of secondary importance.

This translates the region occupied by the models at $1-3~\mathrm{Myr}$ to larger masses, meaning that although all of the discs at high masses with inferred viscous ages $\sim10~\mathrm{Myr}$ are now accessible to the models, the many observed discs with inferred viscous ages $\lesssim 0.3~\mathrm{Myr}$ are once again unexplained. Thus without invoking a scatter in the value of $f_{\rm grow}$, we would shift the problem from high masses to low masses. Moreover, by $5-10~\mathrm{Myr}$, Figure \ref{fig:vary_fgrow}b shows that at a given accretion rate, the models lie at too high masses for nearly all of the discs.
However, we note that there is a degeneracy between the opacity and $f_{\rm grow}$.
If the true opacities are much lower than those we used, such that the observers' equivalent dust masses inferred from the sub-mm flux densities are not overestimates, a larger $f_{\rm grow}$ (i.e. significantly lower dust depletion) would be needed to reasonably replicate the data.

Nevertheless, all of these variations show that enhancing the dust retention by some means in some of the discs could be a viable solution to the issue of the discs in Lupus observed to have high masses, but making such modifications across the whole disc population would not increase the range of masses covered by the models and would pose problems for understanding Upper Sco (with the caveat that we have not included external photoevaporation).

\subsection{Temperature model}
Since our assumptions about the temperature lead to emission from dust at temperatures higher than the $T_{\rm dust} = 20~\mathrm{K}$ assumed in estimating the mass from the sub-mm masses, it is worth examining our sensitivity to this assumption. For a solar mass star, an aspect ratio of $h_0=0.033$ at $1~\mathrm{au}$ corresponds to $T_0\sim279~\mathrm{K}$ at the same distance; now we also consider a cooler disc (by a factor $\sim0.57$) with $h_0=0.025$. This is also included in Figure \ref{fig:vary_dust} as the green track.
We see that the difference between this and the base model is minimal. On timescales $t<t_\nu$, the accretion rate $\dot{M}_{\rm acc} \propto \nu \propto T \propto T_0$ for an $\alpha$ model, so is slightly lower. However in the Rayleigh-Jeans limit $B_\nu(T) \propto T$, so the sub-mm flux and observers' equivalent mass $M_{\rm d,obs} \propto T_0$ also, so the observers' equivalent dust masses are also lower, for a given amount of dust. These temperature dependences cancel when evaluating $t_{\rm acc}$, so the inferred viscous age of the disc is insensitive to our choice of temperature. The temperature cannot therefore be a factor in explaining the discs that appear too old.

\subsection{Stellar Dependence}
The evolution of the disc potentially depends on the central star in a number of ways, including the orbital dynamics, the temperature of the disc and the photoevaporation rate, as well as any dependence that the initial disc parameters have.
The variation of the disc temperature with the stellar mass is not well understood - observationally there appears to be little correlation \citep{Tazzari_2017}, but theoretical radiative transfer models find potentially steeper relationships depending on the assumptions such as the opacities and the stellar evolutionary models used to link the stellar luminosity and mass \citep{Sinclair_2020}.
Here, we choose to assume a fairly maximal temperature dependence $T \propto M_*^{1/3}$ ($h \propto M_*^{-1/3}$, $\nu \propto M_*^{-1/6}$).

\begin{figure*}
    \centering
    \includegraphics[width=\linewidth]{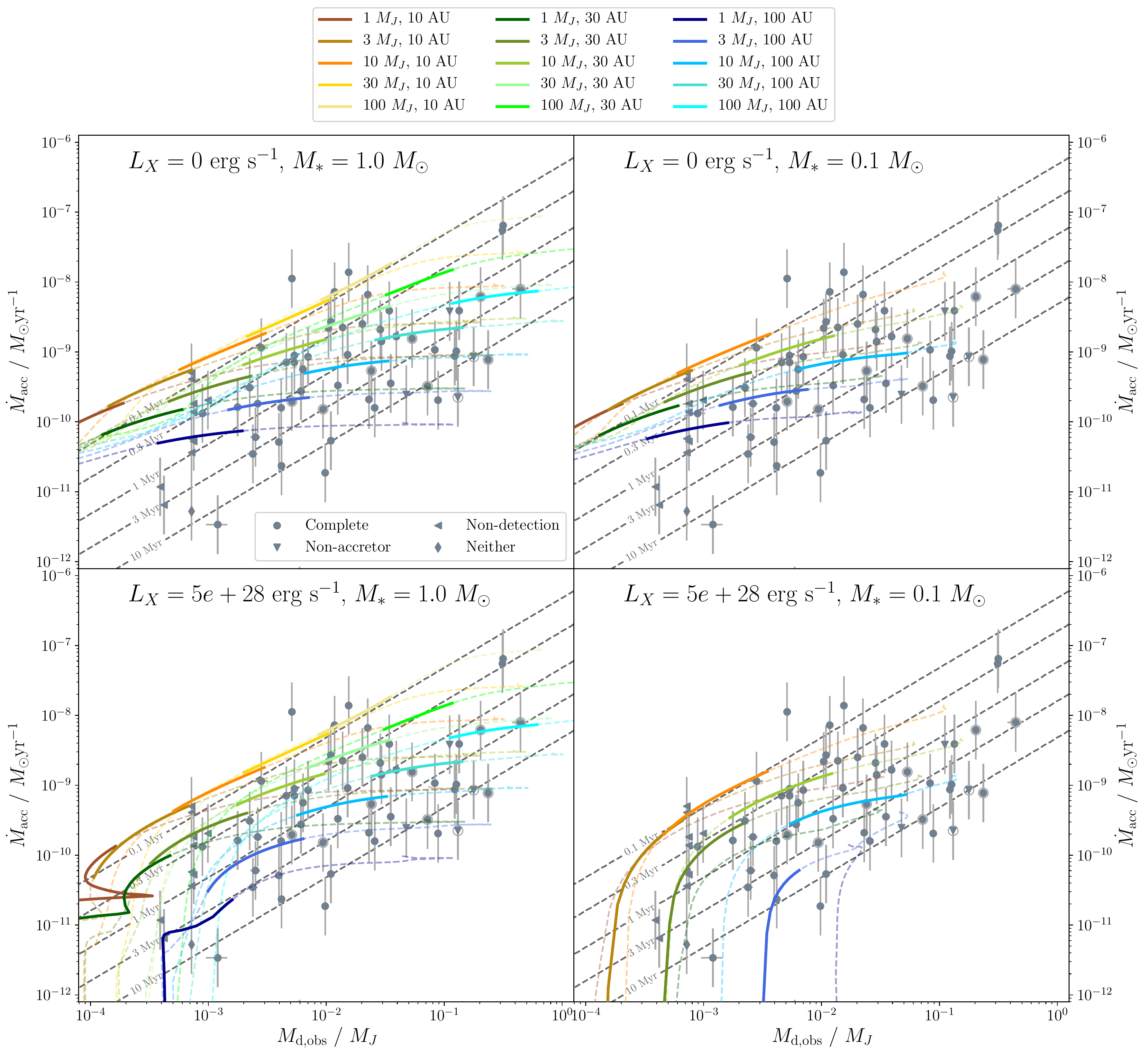}
    \caption{Effect of changing the stellar mass on both a non-photoevaporating model (top) and an X-ray photoevaporation model (bottom). Each panel contains data for the full range of stellar masses from the Lupus catalogue.}
    \label{fig:stellar_mass}
\end{figure*}

The right-hand panels of Figure \ref{fig:stellar_mass} show the impact of repeating the modelling at a mass of $0.1~\mathrm{M_{\sun}}$ (the lowest mass in the \citet{Ansdell_2016} sample) for both the basic model and a model with X-ray luminosity $5\times10^{28}~\mathrm{erg~s^{-1}}$ appropriate to a star of this mass. We limit the initial disc masses $\leq10~\mathrm{M_{\rm J}}$ since a disc-star mass ratio of $0.1$ is typically taken as the limit of gravitational stability. The panels from Figure \ref{fig:4panels_LX} showing the evolution for a solar mass star with the same luminosities are replicated on the left for ease of comparison.

In the non-photoevaporating case, the outcomes are not changed much when considering lower mass stars with only a small preference towards lower masses at a given accretion rate.
Once the dust sizes reach the radial drift limit, the radial drift timescale $t_{rd} \propto M_*^{-1/2}$, hence around lower mass stars radial drift takes longer and is slightly less efficient, but only by a factor $\sim3$ \citep{Pascucci_2016} which is small compared to the overall dust depletion (moreover, this is offset somewhat by the initial radial drift in the fragmentation limited regime being faster for a low mass star).
On the other hand, the discs around low mass stars are cooler and less luminous. Thus, the use of $T_{\rm dust} = 20~\mathrm{K}$ becomes more representative than for solar mass stars and the observers' equivalent dust masses are more accurate determinations of the true dust mass and not so strongly overestimated\footnote{This can be seen in the fact that for a given mass the tracks start at a much lower observers' equivalent dust mass.}.
The size of this effect leading to lower values of the observers' equivalent dust mass for a given true mass is similar to the effect radial drift has on the true masses but in the opposite direction, resulting in very little overall change in the observers' equivalent dust masses at $1-3~\mathrm{Myr}$.

This means that the models retain the inability to reproduce the highest mass discs. However since disc masses are known to be positively correlated with stellar masses \citep{Andrews_2013,Ansdell_2016,Pascucci_2016}, we would not necessarily expect models for a $0.1~\mathrm{M_{\sun}}$ star to replicate the high mass discs.
Therefore as long as the initial disc mass range spans $1-100~\mathrm{M_{\rm J}}$, it doesn't matter precisely how the initial disc masses relate to stellar masses in order to replicate the area of the $\dot{M}_{\rm acc}-M_{\rm dust}$ plane occupied by the data, though the exact distribution in this plane, and the stellar mass dimension may be able to constrain this.

A slightly bigger difference emerges at low accretion rates once photoevaporation is considered. Because of the slower radial drift in the drift limited phase, more dust is retained once the gap opens and the accretion rates decline. At dispersal, this is reflected in both the true dust masses and the observers' equivalent dust masses.
A smaller range of initial parameters can reproduce the sources with disc non-detections, particularly if the upper limits on mm flux found by \citet{Ansdell_2016} turn out to correspond to flux measurements significantly lower than these upper limits (more mild differences would be challenging for the more strongly accreting sources). However, there is no evidence at present that this is the case and that there is any problem; moreover if such a problem emerged as higher sensitivity data becomes available, the relevant sources could simply correspond to stars with below-average X-ray luminosities that disperse later and have more chance to lose dust.

Furthermore, we find that by the time such low stellar masses evolve to the age of Upper Sco, none of these discs would have ongoing accretion, in contrast to the observation of accretion onto stars as low as $0.12~\mathrm{M_{\sun}}$ \citep{Manara_2020} (note when we considered solar mass stars, such as in Figure \ref{fig:euv_old}, all the discs that survived this long had masses $\geq 10~\mathrm{M_{\rm J}}$).
Therefore, the observations for lower mass stars would be better fit by a lower photoevaporation rate, i.e. by a model with $L_{\rm X}<5\times10^{28}~\mathrm{erg~s^{-1}}$, which is not unreasonable since there can be $\sim 0.6~\mathrm{dex}$ scatter about the mean luminosity at a given mass \citep{Preibisch_2005}. Alternatively, the photoevaporation rates at a given X-ray luminosity are calculated in the maximal case of no screening by neutral material close to the star - if such neutral material is present it could give a physical reason why we prefer a slightly lower mass-loss rate.

This preference for a slightly lower mass loss rate could also agree with the fact that $L_{\rm X}=5\times10^{28}~\mathrm{erg~s^{-1}}$ was the best fit to the data assuming a solar mass, despite being at least an order of magnitude below the mean X-ray luminosity at that mass.

\section{Discussion: Companions and Substructure}
\label{sec:other}
\subsection{Binaries}
For the sample of stars observed by \citet{Ansdell_2016}, 4 were known binaries \citep[Sz 68, Sz 74, Sz 81A, and V856 Sco,][]{Ansdell_2018}. Of these, Sz 74 is notable for lying at an accretion rate higher than the regions accessible to the models, though within the typical $0.42~\mathrm{dex}$ \citep{Alcala_2017} error bars.\footnote{Sz 88A, another high accretion rate source, is also known to have a companion at a separation of 1.5 arcseconds \citep{Reipurth_1993}, though \citet{Ansdell_2018} only found a secondary millimetre component at 0.34 arcseconds.}
We now consider whether its high accretion rate could be accounted for by its duplicity.

Observationally, the high accretion rate of Sz 74 could result because Sz 74 does not appear in the list of binaries in \citet{Alcala_2014}. This would have knock-on effects in converting the accretion luminosity to the accretion rate if, for example, the secondary contaminated their estimate of the star's radius (obtained by fitting the dereddened SED to a reference SED).

Physically, a binary star system imposes an outer constraint on each circumstellar disc at the tidal radius $R_t$. This modifies the analytical solution from the similarity solution given in Equation \ref{eq:LBPprofile} such that the disc mass and accretion rate decline exponentially \citep{Rosotti_2018}
\begin{equation}
    M_{\rm disc} = M_{\rm disc,0} \exp\left( - \frac{t}{t_\nu(R_t)} \right)
    ,
\end{equation}
where $t_\nu(R_t) = \frac{16}{3\pi^2} \frac{R_t^2}{\nu(R_t)}$.
This fixed outer boundary (as opposed to the spreading truncation radius $R_{\rm out} = R_{\rm C} (1+t/t\nu)$ of Equation \ref{eq:LBPprofile}) means that the accretion timescale is constant, preventing the disc from ageing in the $\dot{M}_{\rm acc}-M_{\rm disc}$ plane:
\begin{equation}
    t_{\rm acc} = \frac{M_{\rm disc}}{\dot{M}_{\rm acc}} = t_\nu(R_t)
    \label{eq:tacc_binary}
    .
\end{equation}

\citet{Ansdell_2018} detect mm emission from the known secondary component of Sz 74 (HN Lup) at a separation 0.31 arcseconds (in agreement with previous measurements), corresponding to a separation of $46.5~\mathrm{au}$ at $150~\mathrm{pc}$. The binary has a mass ratio consistent with $1$ \citep{Woitas_2001}, making the tidal radius $15.5~\mathrm{au}$.
Using Equation \ref{eq:tacc_binary} with our standard parameters, this small tidal radius would correspond to $t_{\rm acc} \approx 1.2~\mathrm{Myr}$, compared to $t_{\rm acc} > 2~\mathrm{Myr}$ for a cluster of single discs with ages $\geq 1~\mathrm{Myr}$.
Therefore we should indeed expect Sz 74 to be a relative outlier that cannot be well explained by the models for discs around single stars (even if we account for dust, as the depletion, and its effect on $t_{\rm acc}$, should be similar to the most compact discs around single stars in the sample).

\subsection{Dust Trapping in Substructures}
Save for where a gap is opened by photoevaporation, our model neglects the possibility of disc substructures. These are thought to be fairly common, and some degree of structure was ubiquitous in the DSHARP sample \citep{Andrews_2018,Huang_2018}, though this was biased towards bright - and therefore massive - discs around bright sources in order to get sufficient contrast on the features. The most tantalising, and popular, explanation for the annular features is that they result from the presence of planets \citep[e.g.][]{Papaloizou_1984,Paardekooper_2004,Zhang_2018}. \citet{Sinclair_2020} suggest however that it becomes harder for planets to open gaps around lower mass hosts, which may challenge ubiquity across the stellar mass - and disc mass - range.

Of particular interest here are discs with large dust rings as these are consistent with the trapping of large dust grains in pressure maxima \citep{Dullemond_2018}, thus allowing dust grains to survive for much longer \citep{Pinilla_2012}. In the DSHARP sample, the highest contrast ring in Lupus is found for GW Lup (Sz 71), which is one of the sources with an inferred viscous age too old to fit the basic dust model. Other discs that cannot be fit by the dust model also show azimuthal substructures at lower-contrast. Since the contrast is affected by the convolution with the ALMA beam \citep{Huang_2018}, it is less clear whether dust traps would be effective in these sources, though \citet{Pinilla_2012} suggest a variation of amplitude 30 per cent in the gas surface density would be sufficient to trap the dust. Brightness reconstructions which achieve sub-beam resolution \citep[e.g.][]{Jennings_2020} could help more accurately constrain the width and amplitude of these traps.
\citet{Long_2020} recently showed that there are substructures at sub-beam scales in GQ Lup, a much more compact disc than any of the DSHARP sample, though still a relatively high mass one.

Moreover, several of the sources with large inferred viscous ages $\sim10~\mathrm{Myr}$ have been identified as transition discs; \citet{Pinilla_2018} show that the slope of the relation between $M_{\rm dust}$ and $M_*$ is much shallower for transition discs than full discs. Since \cite{Pascucci_2016} found radial drift was necessary to explain the relation for full discs, \citet{Pinilla_2018} concluded this may be evidence that dust trapping in transition discs is impeding radial drift and having the effect of keeping their observed masses high; more recent dust evolution modelling by \citet{Pinilla_2020} showed that the shallow $M_{\rm dust}-M_*$ relation for transition discs and discs with substructures could be explained by pressure traps but would require inhibition of boulder formation within the traps.

In the older sample from Upper Scorpius, there are only two discs at much too high mass to be explained by the photoevaporating models
\citep[one of which is HD 143006, another source with high-contrast dust rings that could be explained by trapping,][]{Perez_2018,Dullemond_2018}.
This is relatively few compared to Lupus and could be evidence for the influence of substructures in promoting disc clearing: \citet{Rosotti_2013} found that a giant planet would reduce the accretion flow in the inner disc allowing photoevaporation to clear the disc sooner than otherwise.
Thus if substructure is more common for high mass discs (the only ones where it has been directly observed, and the ones for which our dust models do not match) then massive substructure-bearing discs could have shorter lifetimes than lower-mass, potentially structureless, counterparts, with only the latter surviving to the age range of Upper Sco. In contrast, a deceleration of the dust growth as explored in the previous section (see Figure \ref{fig:vary_fgrow}b) would predict higher mass discs persisting to $5-10~\mathrm{Myr}$.

Therefore dust retention in substructures could be responsible for keeping the measured disc masses high, and is thus a promising explanation for several of the discs in Lupus with large masses for their ages.
However, in this work we find the majority of the discs in the datasets can be explained without dust trapping, so substructures need not be ubiquitous across the disc mass range.

\section{Conclusions}
\label{sec:conclusions}
We used a dust evolution model to follow the evolution of protoplanetary discs for a range of initial parameters. We translated the resulting millimetre fluxes predicted by the models into the corresponding values of dust mass that an observer would deduce from the millimetre flux (which we term "observers' equivalent dust mass") using standard assumptions about the dust-to-gas ratio, opacity and temperature distribution in the disc and compared these to estimates of disc mass from sub-mm fluxes and accretion rates from the UV continuum excess. We then considered the predictions of the model for the evolution of discs in the plane of accretion rate versus apparent disc mass and compared these tracks with observational datasets in the Lupus and Upper Sco star forming regions.

We find that when the growth and radial drift of dust (which can lower the dust-to-gas ratio by over an order of magnitude, see lower left panel of Figure \ref{fig:4panels_nope}) are taken into account, viscous disc models do a good job at explaining the distribution of sources in the plane of accretion rate versus (observers' equivalent) dust mass. Viscous models predict that the true disc mass and accretion rate should be related (in the case $\nu \propto R$) such that $t_{\rm acc}= M_{\rm disc}/\dot{M}_{\rm acc} > 2 t$ where $t$ is the age of the disc. Previously it has been puzzling that many sources have $t_{\rm acc}$ values that are much less than the ages of the star forming region.
We show that this can be explained if, due to a high gas-to-dust ratio, the true disc mass is significantly larger than what has been conventionally inferred from sub-mm measurements and demonstrate that the use of dust evolution models can largely achieve consistency between viscous models and observations:

\begin{enumerate}
    \item Since the dust masses at a given accretion rate are lowered through radial drift, the models occupy the region of apparently low $t_{\rm acc}$ (`Observational' panels of Figure \ref{fig:4panels_nope}, Figure \ref{fig:nope_old}) - which is not reproduced when considering the gas masses (`Theoretical' panels of Figure \ref{fig:4panels_nope}, Figure \ref{fig:nope_old}) - at an age broadly compatible with the estimated ages of the star forming regions.
    \item Since the efficiency of radial drift depends strongly on the initial disc radius, the scatter in accretion rates at a given (observers' equivalent) dust mass is increased versus that predicted by the gas masses, and is more consistent with the observations.
    \item To achieve sufficient scatter in the accretion rates, we find that we prefer low viscosities with $\alpha \lesssim 10^{-3}$.
    \item There is an upper boundary to the accretion rates at a given dust mass that, unlike the prediction ($t_{\rm acc} > 2 t$), does not evolve with age and is due to an interplay between the dust abundance and the opacity and temperature of the dust. This boundary at $t_{\rm acc} \sim 0.1~\mathrm{Myr}$ is consistent with the datasets from Lupus and Upper Sco wherein a similar lack of evolution with age was noted by \citep{Manara_2020}.
    \item Despite the dust-to-gas ratio $\epsilon$ deviating significantly from $0.01$, this does not eliminate the correlation between accretion rates and masses expected from viscous theory.
    Thus, while the correlation is not evidence for the validity of assuming $\epsilon=0.01$ as argued by \citet{Manara_2016}, and the true disc masses may be significantly larger than typically inferred, the dust masses are likely better indicators of the total mass than measurements of CO isotopologues.
\end{enumerate}

Despite these successes, some discs (mainly those in Lupus with $t_{\rm acc} \sim 10~\mathrm{Myr}$) are observed to have high apparent dust masses, despite the models evolving through that region very quickly. In Figures \ref{fig:vary_dust}, \ref{fig:vary_alpha} \& \ref{fig:vary_fgrow} we showed how altering the dust fragmentation velocity ($u_f$), dust growth timescale ($f_{\rm grow}$), or viscous $\alpha$ parameter could reduce the efficiency of radial drift. However, any solution applied uniformly to all discs may create further discrepancies between our models and the data.
Alternatively, as discussed in Section \ref{sec:other}, annular substructures, as known from high resolution disc observations, could be acting as efficient traps of dust in these discs. 

An additional discrepancy is the population of discs detected in the sub-mm with dust masses of $\sim10^{-3}~\mathrm{M_{\rm J}}$ but very low accretion rates $<10^{-10}~\mathrm{M_{\sun}}~\mathrm{yr^{-1}}$. Because radial drift acts rapidly compared with the viscous evolution, in viscous models the accretion rate should not decline to such a low value until the dust emission has declined to unobservably low values. In Section \ref{sec:photoevaporation} we show that photoevaporation may help, with the following conclusions:

\begin{enumerate}
    \item Internal photoevaporation acts to starve accretion onto the star, ultimately by opening a gap in the disc; by cutting off accretion before the dust has all drifted on to the star (with any remaining dust becoming trapped in the outer disc), photoevaporation resolves this discrepancy (Figures \ref{fig:4panels_euv}, \ref{fig:4panels_LX} \& \ref{fig:euv_old}). Relatively low photoevaporation rates $\lesssim10^{-9}~\mathrm{M_{\sun}}~\mathrm{yr^{-1}}$ - as appropriate to an EUV driven model with $\Phi=10^{42}~\mathrm{s}^{-1}$ or an X-ray model with $L_{\rm X}=5\times10^{28}\mathrm{erg~s^{-1}}$ - are required to reproduce the observed dust masses as discs pass through these low accretion rates (Figure \ref{fig:4panels_LX}).
    \item Despite their ability to constrain the mass loss \textit{rates}, the data do not discriminate between models where the photoevaporation is EUV or X-ray driven.
    \item If at all times, some discs are approaching or undergoing rapid dispersal, then we should expect to see accretion rates roughly extending to a constant floor scale where they equal the photoevaporation rate. In combination with the above argument for the upper limit, this explains the lack of evolution in the range of accretion rates between Lupus and Upper Sco noted by \citep{Manara_2020}. 
    \item In order to further constrain the photoevaporation rates, high sensitivity sub-mm observations of the low mass discs which currently lack statistically significant detections - and in Upper Sco, accretion rates for these sources - would be useful.
\end{enumerate}

\section*{Acknowledgements}
We thank the anonymous reviewer for a very constructive report which greatly helped improve the quality of our modelling and data presentation.
We are grateful to Giovanni Rosotti for useful discussions on this work.
ADS thanks the Science and Technology Facilities Council (STFC) for a Ph.D. studentship and RAB and CJC acknowledge support from the STFC consolidated grant ST/S000623/1.
This work has also been supported by the European Union's Horizon 2020 research and innovation programme under the Marie Sklodowska-Curie grant agreement No 823823 (DUSTBUSTERS).

\section*{Data Availability}
The observational datasets were derived from sources in the public domain using the VizieR catalogue access tool:
https://dx.doi.org/10.26093/cds/vizier.18280046,
https://dx.doi.org/10.26093/cds/vizier.35610002,
https://dx.doi.org/10.26093/cds/vizier.36000020; and from https://doi.org/10.1051/0004-6361/202037949.

The code from which model data were generated is available from Zenodo at https://doi.org/10.5281/zenodo.3981378.




\bibliographystyle{mnras}
\bibliography{accretion_mass_relation} 




\bsp	
\label{lastpage}
\end{document}